# Reactivity screening of single atoms on modified graphene surface – From formation and scaling relations to catalytic activity

Aleksandar Z. Jovanović[1], Slavko V. Mentus[1,2], Natalia V. Skorodumova[3,4], Igor A. Pašti[1,3,*]

[1]*University of Belgrade – Faculty of Physical Chemistry, Belgrade, Serbia*

[2]*Serbian Academy of Sciences and Arts, Belgrade, Serbia*

[3]*Department of Materials Science and Engineering, School of Industrial Engineering and Management, KTH – Royal Institute of Technology, Stockholm, Sweden*

[4]*Department of Physics and Astronomy, Uppsala University, Uppsala, Sweden*

**Abstract**

Single atom catalysts (SACs) present the ultimate level of catalyst utilization, which puts them in the focus of current research. For this reason, their understanding is crucial for the development of new efficient catalytic systems. Using Density Functional Theory calculations, model SACs consisted of nine metals (Ni, Cu, Ru, Rh, Pd, Ag, Ir, Pt and Au) on four different supports (pristine graphene, N- and B-doped graphene and graphene with single vacancy) were analyzed. Among them, only graphene with a single vacancy enables the formation of SACs, which are stable in terms of aggregation and dissolution under harsh conditions of electrocatalysis. Reactivity of models SACs was probed using atomic (hydrogen and A = C, N, O and S) and molecular adsorbates ($AH_x$, $x$ = 1, 2, 3 or 4, depending on A), giving nearly 600 different systems included in this study. Scaling relations between adsorption energies of A and AH$_x$ on model SACs were confirmed. However, the scaling is broken for the case of CH$_3$. There is also an evident scaling between adsorption energies of atomic and molecular adsorbates on metals SAs supported by pristine, N-doped and B-doped graphene, which originates from similar electronic structures of SAs on these supports. Using the obtained data, we have analyzed the hydrogen evolution on the model SACs. Only M@graphene vacancy systems (excluding Ag and Au) are stable under hydrogen evolution conditions in highly acidic solutions. Additional interfacial effects are discussed and the need for proper theoretical treatment when studying SACs interactions with molecular species.

---

[*] **Corresponding author:** e-mail: igor@ffh.bg.ac.rs

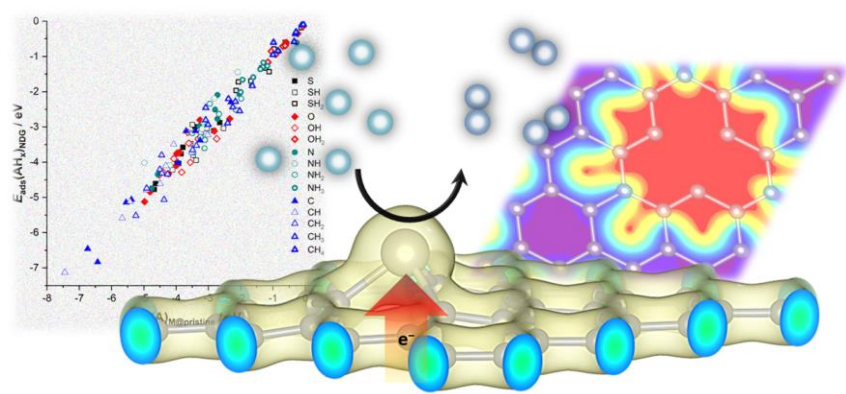

## 1. Introduction

Battling global challenges related to the energy requirements of our society requires tremendous efforts and innovative approaches. Many processes we use these days for energy conversion or storage require catalysts, and an increasing number of them actually rely on electrocatalysis.[1] It is a fact that electrocatalytic processes most frequently use expensive Pt-group metals (PGM). This makes them less attractive for large-scale use. Strategies to overcome this problem are (i) finding new PGM-free electrocatalysts or (ii) reducing the content/loading of PGM. Considering the latter point, the first step is reducing size while increasing the number of active sites. However, even at the nanoscale, a nanoparticle's interior is catalytically inactive and presents a clear loss. For this reason, core-shell or thin-film catalysts have been developed.[2,3] However, the absolute extremes of the catalyst utilization are Single Atom Catalysts (SACs). In SACs where every single atom has the catalytic function.[4,5] In this case, the catalyst support is vital. As every single atom is in contact with the support, there is a strong interplay between them.[6,7] For this reason, the same metal atom at two different supports can behave entirely differently. In a way, this is an advantage as the phase space of SACs is enormous. However, it is also a challenge, as we need to understand this interplay to develop new SACs rationally.

The field of SACs developed tremendously in the last decade. Now we have many different SACs for various (electro)catalytic reactions. Only a few examples are the review of heterogeneous SACs,[8-10] the review of Single Atom Alloy catalysts as a class of SACs[11] and others. Different aspects of SACs have been summarized in many timely reviews, like design principles and applications,[12] application of SACs in organic chemistry,[13] heterogeneous SACs for $CO_2$ electroreduction,[14] SACs for electrocatalytic hydrogen production,[15] synthetic strategies and electrochemical applications of SACs,[16] SACs for the green synthesis of fine chemicals,[17] SA catalysis on carbon-based materials[18] and others, addressing some general features of SACs or summarizing SACs for a given (electro)catalytic reaction(s).

Many individual reports on SACs contain theoretical contributions, most frequently within Density Functional Theory (DFT). While nanoparticles cannot be routinely analyzed using DFT, SACs, on the other hand, present perfect study cases. The structure of a SAC can be transferred into the DFT model with much fewer approximations. For example – nanoparticles are most frequently modeled part-by-part, using single-crystal surfaces of a given composition and orientation. In this way, one can obtain an in-depth, electronic structure-level description of a SAC considered for a given reaction. However, we firmly believe that, in the case of SACs, the real power of DFT lies in the possibility of investigating a large number of systems and *understanding trends*.[19]

In this report, we present a detailed study of model SACs consisted of 9 transition metal atoms (Au, Ag, Cu, Ir, Ni, Pd, Pt, Rh, Ru) and graphene-based surfaces (pristine graphene, N-doped and B-doped graphene, and graphene with single vacancy). The metals are chosen as being frequently is in catalytic processes. Graphene-based surfaces were selected due to high conductivity and large surface area, making them attractive for building SACs for (electro)catalysis. In order to understand the overall trends, we have analyzed (i) formation of SACs and electronic structures, (ii) mobility of single atoms on the considered supports, (iii) stability under electrochemical conditions, (iv) screened reactivity of model SACs for different adsorbates (H, C, N, O and S and corresponding $AH_x$ molecules; $x$ = 4 for C, 3 for N, and 2 for O and S), (iv) analyzed hydrogen evolution reaction (HER) activity trends on model SACs, and (v) addressed some computational issues related to theoretical modeling of SACs.

## 2. Methods

The adsorption of 9 transition metal atoms (Au, Ag, Cu, Ir, Ni, Pd, Pt, Rh, Ru) was investigated on the (modified) graphene surfaces in a 4×4 supercell (32 atoms). Pristine, graphene with a single vacancy (VG), B-doped (BDG), and N-doped graphene (NDG) were considered (**Figure 1a**). Doping was performed substitutionally, so the doped structure had the formula of $C_{31}X$, X being B or N.

Calculations were based on DFT using the Generalized Gradient Approximation in the PBE parametrization.[20] The calculations were performed with Quantum ESPRESSO *ab initio* package (QE), using ultrasoft pseudopotentials.[21] Periodic graphene sheets were separated from each other by 15 Å of vacuum. The kinetic energy cutoff of the plane-wave basis set was 490 eV, and the charge density cutoff was 16 times higher. Spin polarization was included for all the investigated systems. The contribution of the long-range dispersion interactions was investigated using the DFT+D2 scheme by Grimme, as implemented in QE.[22] For certain systems, calculations were also done without the D2 correction. A Monkhorst-Pack $\Gamma$-centered 4×4×1 *k*-point mesh was used for structural optimization.[23] DOS calculations were done using denser 20×20×1 *k*-point mesh.

Three adsorption sites were considered on pristine graphene: T, B, and H (**Figure 1b**). Upon introducing the vacancy or dopants, the symmetry of the supercell is lowered, resulting in 25 nonequivalent adsorption sites (**Figure 2c**). These sites were later assigned to every top, bridge, and hollow site within the supercell using simple symmetry operations. Boron and nitrogen were selected for doping since they induced minimal structural change upon substitutional doping, remaining in the graphene basal plane.[24]

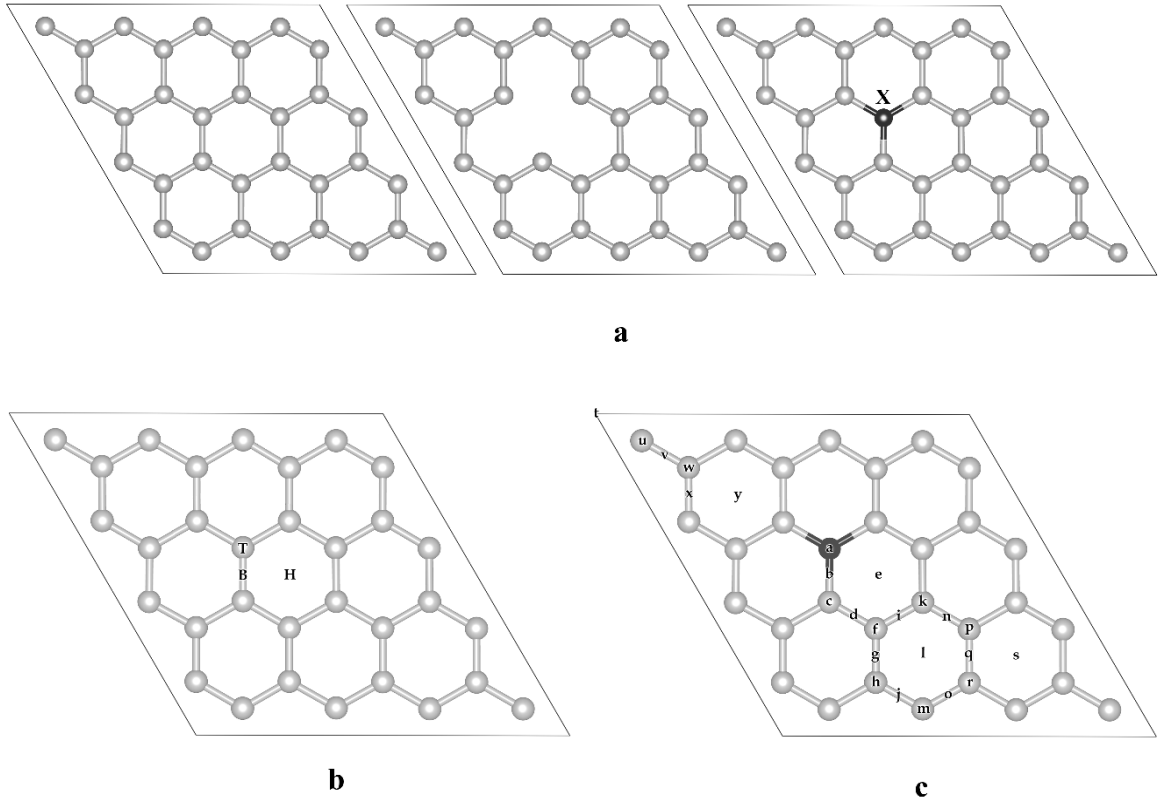

**Figure 1.** (a) Pristine, monovacant, and doped graphene models, X denotes the dopant (B or N) position. Bottom - Naming scheme for adsorption sites on pristine (b) and modified graphene (c).

The graphene lattice was fully optimized with the given calculation parameters. The C–C bond length and the value of cohesive energy were calculated to be 1.42 Å and −7.84 eV, respectively, similar to previously reported values.[24–26] Vacancy formation energy was calculated as:

$$E_{vf} = E_0^{vac} - \frac{n-1}{n} E_0^{G} \qquad \text{[Equation 1]}$$

Where $E_0^{vac}$ and $E_0^{G}$ are the ground state energies of the monovacant and pristine graphene systems, and n is the total number of atoms in the supercell ($n$ = 32). The calculated value of 7.76 eV is in good agreement with previous reports.[27,28]

Doping with B and N was performed substitutionally on the vacancy site, and the dopant binding energy was calculated as:

$$E_{dop} = E_0^{V+X} - E_0^{vac} - E_0^{X} \quad \text{[Equation 2]}$$

Where $E_0^{V+X}$ and $E_0^{X}$ are the ground state energies of doped graphene and the isolated dopant atom, respectively, with X denoting B or N. Dopant binding energies were calculated to be −13.15 eV for B-doped, and −11.74 eV for N-doped graphene, in close agreement with our previous results.[24] Magnetic behavior is observed only in the case of monovacant graphene, the vacancy giving rise to a total magnetic moment of 1.74 $\mu_B$, somewhat higher than previously observed, and close to the value calculated by Rodrigo *et al.* for a 30×30 cell.[29]

Metal ad-atom adsorption was probed at each of the sites mentioned above. During structural optimization, the relaxation of all atoms in the simulation cell was allowed. The adsorption energy of single metal atoms ($E_{ads}(M)$) was evaluated as:

$$E_{ads}(M) = E_{M@G} - E_G - E_M \quad \text{[Equation 3]}$$

where $E_{M@G}$, $E_G$ and $E_M$ stand for the total energy of (modified) graphene surface with adsorbed metal M, the total energy of the corresponding clean surface and the total energy of the isolated metal atom, respectively. The adsorption energy of the adatom at the site it finally relaxed at was ascribed to the initial site in which it was placed (e.g., in Figure 2: *c*, *d*, *f*, and *y* sites all relaxed into the *b* site; thus, the adsorption energy of the *b* site was taken for all four initial sites. Based on this approach, heatmaps (Figure 2, right), which show reactivity of the entire 4×4 supercell, were constructed for each surface and each metal under the study.

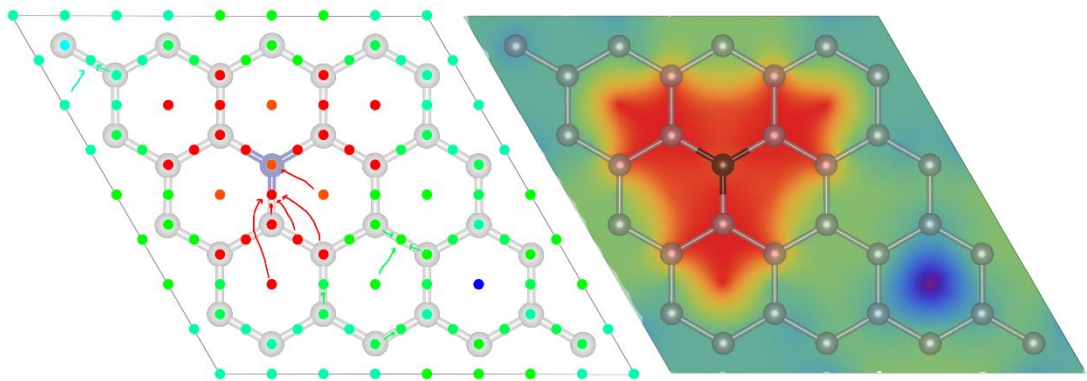

**Figure 2.** Adsorption energies of discrete sites plotted as color-mapped dots (left), and as a heatmap upon linear interpolation of individual values (right). The direction of the arrows on the left shows the relaxation from the initial to the preferred site. Adsorption of Pt on B-doped graphene is used as an example. Values in red represent stronger, while the values in blue represent a weaker bonding.

Once the preferential adsorption/tethering sites for single atoms were determined, we investigated the atomic and molecular adsorption on these model SACs. The adsorption of H, C, N, O and S and corresponding $AH_x$ molecules ($x = 4$ for C, 3 for N, and 2 for O and S) was analyzed. Adsorption was quantified using the adsorption energy ($E_{ads}(A)$ or $E_{ads}(AH_x)$), defined as:

$$E_{ads}(A) = E_{M@G+A} - E_{M@G} - E_A \qquad \text{[Equation 4]}$$

where $E_A$ stands for the total energy of isolated adsorbate. Following definitions of $E_{ads}(M)$ and $E_{ads}(A)$, stronger adsorption gives more negative values of adsorption energy.

## 3. Results and discussion
### 3.1. Formation of single atom-graphene systems and the viability of model SACs

As the initial step of forming single-atom graphene systems, we consider the deposition of selected atoms on the pristine and modified graphene surfaces. An overview of adsorption parameters is given in Table 1.

**Table 1.** Preferred adsorption sites and adsorption energy of metal atoms on pristine, doped (NDG and BDG), and graphene with a single vacancy (VG)

| | | | | | | | | | | | | |
|---|---|---|---|---|---|---|---|---|---|---|---|---|
| | | | | | | | $E_{ads}(M)$ / eV | | | | | |
| | | pristine | | | BDG | | | NDG | | | VG | |
| M | site* | PBE | PBE+D2 | site | PBE | PBE+D2 | site | PBE | PBE+D2 | site | PBE | PBE+D2 |
| Ni | H | −1.44 | −1.71 | e | −2.18 | −2.47 | y | −1.40 | −1.68 | a | −6.52 | −6.78 |
| Cu | T | −0.26 | −0.53 | a | −1.38 | −1.68 | c | −0.47 | −0.73 | a | −3.63 | −3.89 |
| Ru | H | −1.59 | −1.96 | e | −2.44 | −2.83 | y | −1.92 | −2.29 | a | −8.30 | −8.66 |
| Rh | H | −1.43 | −1.80 | e | −2.30 | −2.67 | y | −1.86 | −2.22 | a | −7.68 | −8.04 |
| Pd | B | −1.08 | −1.49 | b | −1.58 | −1.95 | n | −1.10 | −1.47 | a | −5.11 | −5.46 |
| Ag | H | −0.03 | −0.32 | a | −0.90 | −1.26 | y | −0.04 | −0.33 | a | −1.65 | −2.04 |

| | | | | | | | | | | | |
|---|---|---|---|---|---|---|---|---|---|---|---|
| Ir | B | −1.14 | −1.70 | e | −2.16 | −2.76 | f | −1.69 | −2.26 | a | −8.92 | −9.42 |
| Pt | B | −1.45 | −2.01 | b | −2.12 | −2.68 | d | −1.81 | −2.38 | a | −6.91 | −7.41 |
| Au | T | −0.14 | −0.70 | a | −1.12 | −1.69 | c | −0.72 | −1.29 | a | −2.30 | −2.97 |

*site - preferred adsorption site. Top, bottom, or hollow (t, b, h) for p-graphene; for modified graphene, refer to Figure 1.

The adsorption energies and preferred binding sites on pristine and monovacant graphene are in good agreement with our previous studies of monoatomic adsorption on pristine and monovacant graphene[24,30], as well as studies of various TM atoms adsorption on pristine graphene.[31,32] On pristine graphene, adsorption is preferential on the bridge and hollow sites, except for Au and Cu. Binding on the monovacant graphene is always on the vacancy site. On average, monovacant graphene shows the strongest bonding of TM atoms, followed by B-doped, N-doped, and finally pristine graphene. Dispersive interaction gives a similar total contribution to adsorption energy in most of the cases. However, the relative contribution is most pronounced in weakly binding systems, such as metal-pristine graphene, which is known to be rather chemically inert. From the DOS plots in Figures 3 and 4 (the cases of pristine and VDG, for NDG and BDG, see Supplementary information, Figure S1 and S2), one can observe that the electronic structure of pristine graphene is weakly affected upon metal SA adsorption, in line with small adsorption energies. SA bands are narrow and relatively close to the Fermi level. An almost identical situation is seen for the cases of NDG and BDG.

In contrast, SA embedding in single vacancy results in strong interactions with the graphene matrix and the SA *d*-states are wide and buried well below the Fermi level. Bandgap opening was not observed in any of the studied cases, suggesting that conductivity is preserved (see Figure S1-S4). This is of crucial importance for the electrocatalytic applications.

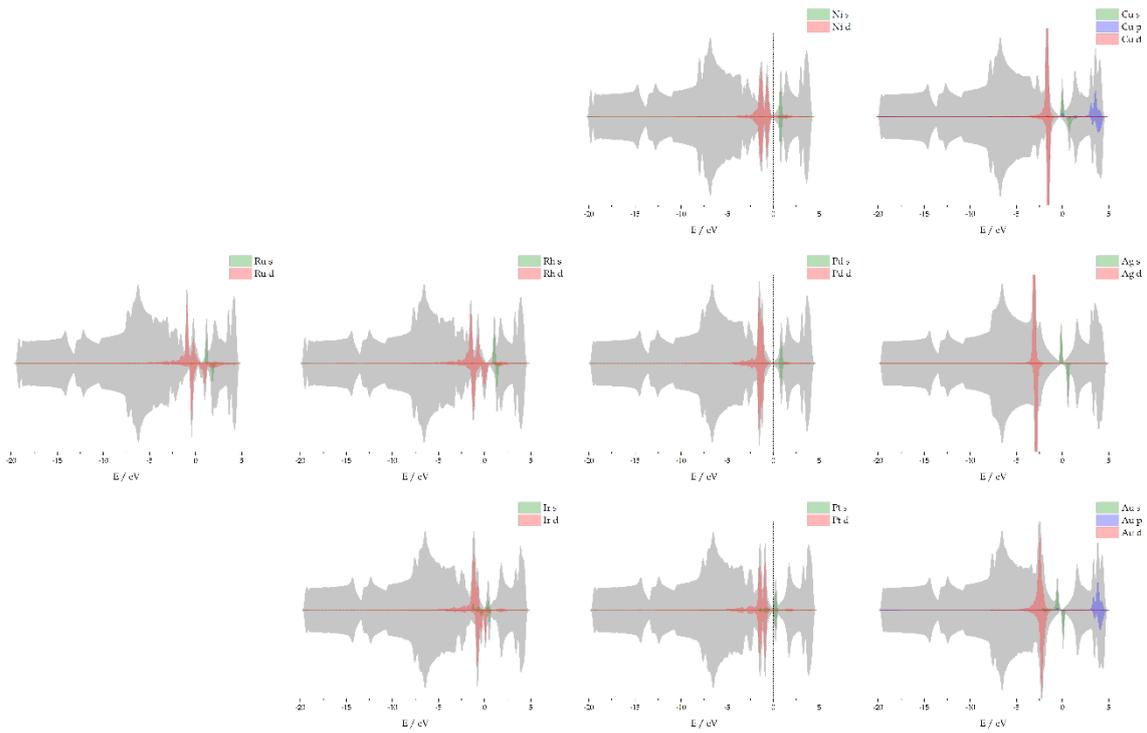

**Figure 3.** DOS plots for single atoms adsorbed on pristine graphene. The energy scale is referred to the Fermi level (vertical dashed line).

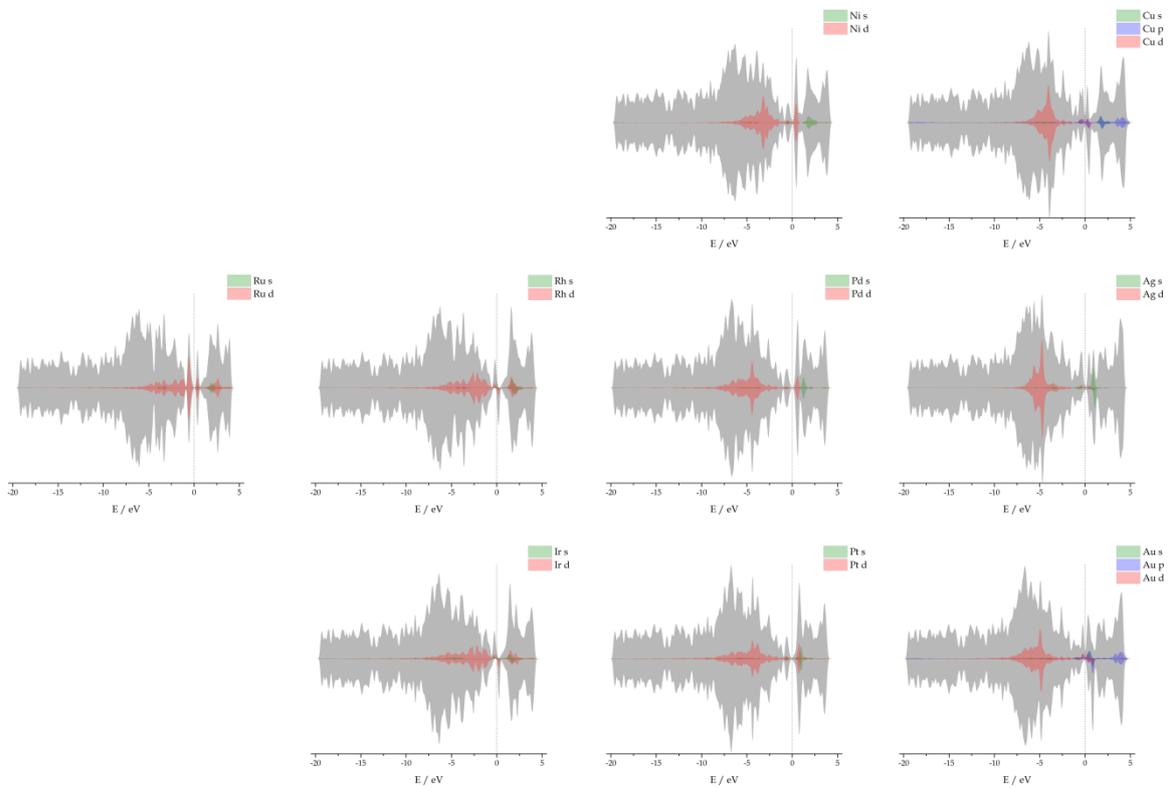

**Figure 4.** DOS plots for single atoms adsorbed on graphene with a single vacancy. The energy scale is referred to the Fermi level (vertical dashed line).

An aspect of the investigated model SACs related to practical application is maintaining SA on the support that is to prevent SA aggregation into nanoparticles. This issue has to be considered from the point of the thermodynamic tendency of SA to form a metal lattice. Also, the ability of SA to migrate over the support in order to aggregate. Point (i) can be quickly checked by comparing $E_{ads}(M)$ and cohesive energy of the corresponding metal ($E_{coh}(M)$), like in our previous work.[24] Such a comparison is presented in Figure 5. As can be seen, among the investigated systems, there are only a few of them from the M@VG family where SA adsorption energy is high enough to overcome cohesive energy and prevent the thermodynamic tendency of metal SA to aggregate.

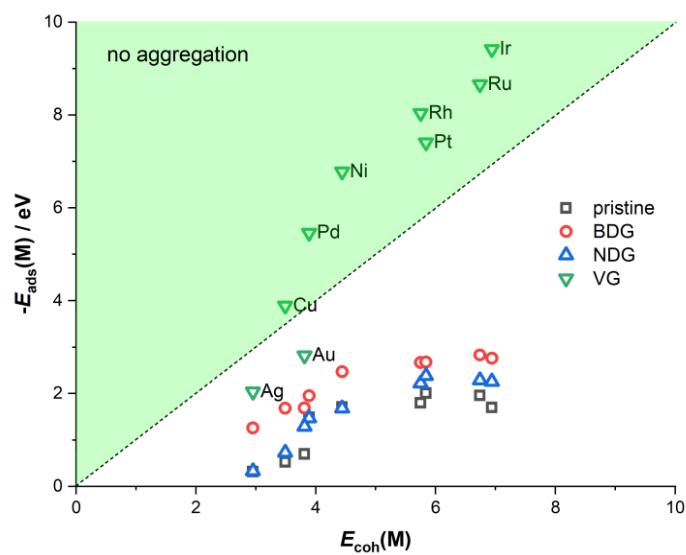

**Figure 5.** Comparison between metal SA adsorption energies on studied graphene surfaces and cohesive energies of clean metal. The systems falling in the shaded area should not show a tendency towards aggregation. Cohesive energies were taken from reference [33].

However, in connection with the point (ii) above, if metal SA is deposited at very low concentration and its mobility is low enough, the probability of segregation of the metal phase will be low. Hence, mobility also must be considered. From the values in Table 1 and the heatmaps of adsorption energy (Figure 6 and Figure S5 and S6), we see that in the case of monovacant graphene, the bonding is on the vacancy site, which presents the anchoring site for all the studied SAs. The presence of a single vacancy influences a large portion of the simulation cell, with SAs relaxing to the vacancy even from the very edges of the supercell. In the case of B-doped graphene, the B dopant itself and the nearest bridge and hollow site are the most active. With N-doped graphene, we observe that the metal SAs bind preferentially on the graphene lattice, either at the top site closest to the N dopant or at the various distant bridge and hollow sites. Considering the overall trends, a single vacancy unambiguously presents a strong anchoring site for SAs. Following the obtained results, we conclude that all studied SAs require at least 1 eV to overcome the barrier and exit the vacancy site. This barrier cannot be overcome at temperatures close to room temperature, suggesting that, in the case of VG, SAs should be stable.

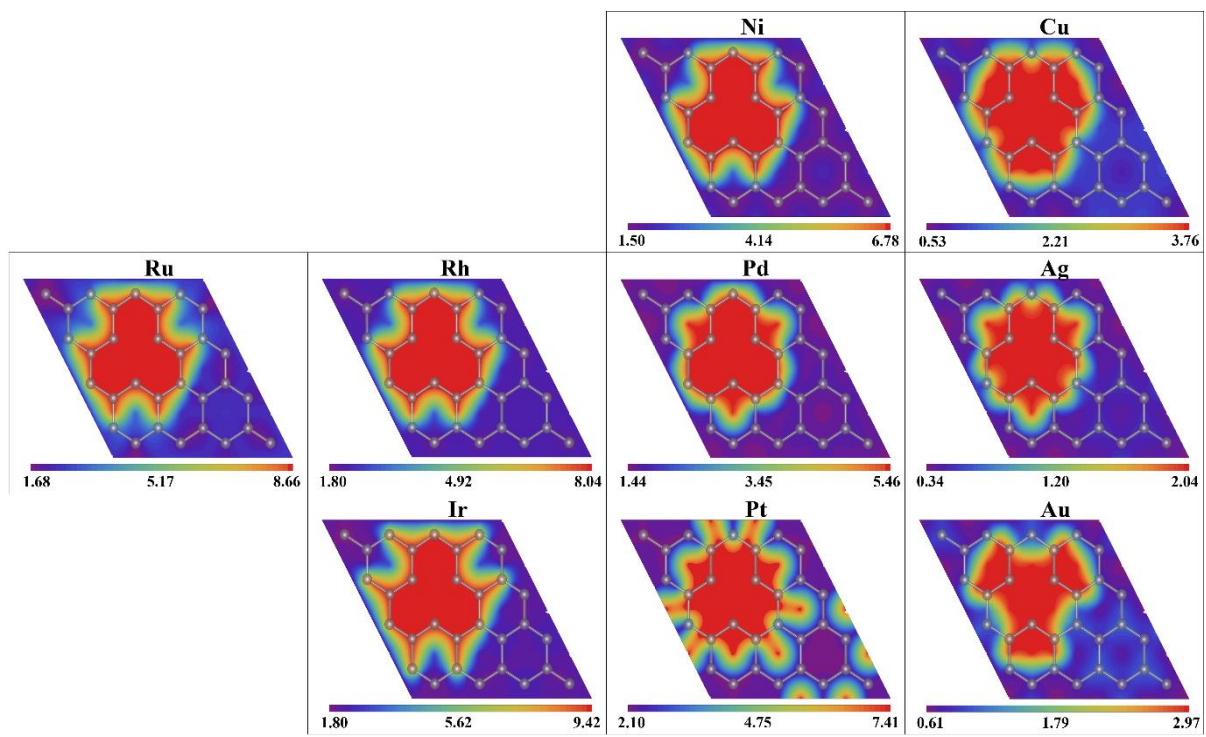

**Figure 6.** Heatmaps of $E_{ads}$(M) in units of eV on VG. The values in red represent stronger, while the values blue represent a weaker bonding. Note that the scale is not the same for different systems.

In the case of BDG, the dopant site (or adjacent carbon atoms) also presents the anchoring site for metal SA. The energy surface for SA diffusion is much smoother compared to the VG case, but still, at least 0.5 eV is required for SAs to migrate away from the anchoring site, which is still appreciable activation energy for room temperature conditions. Hence, if the concentration of SAs is low and comparable to the concentration of B atoms embedded in the graphene lattice, it should be possible to make SACs using BDG as SA support, although there is a strong tendency for SA to aggregate (Figure 5). In the case of NDG, the energy landscape for diffusion is smooth, while SAs avoid quaternary nitrogen sites. In combination with the tendency of SAs to aggregate on NDG, it can be concluded that graphene with substitutional nitrogen is not the right candidate for SAs support. SAs would easily migrate over NDG and form larger aggregates.

So far, the SAC|vacuum interface was analyzed. However, these days SACs find their place in electrocatalysis as well. Hence, it is essential to address SA metals' tendency to dissolve from the support, as well, as electrocatalysis typically uses harsh acidic or alkaline solutions. As a great majority of electrocatalytic reactions occur in aqueous solutions, we address the thermodynamic stability of metal SA concerning the theoretical electrochemical window in water, limited by hydrogen evolution reaction (HER) on the cathodic side and oxygen evolution reaction (OER) on the anodic side. The reaction considered is

$$M^{z+} + \text{support} + ze^- \rightarrow M@\text{support}.$$

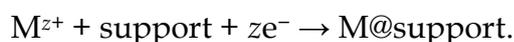

For this purpose, we used the approach as in our previous paper.[24] In brief, the standard electrode potential for the above half-reaction (*vs*. SHE) is obtained by correcting the standard electrode potential for the half-reaction

$$M^{z+} + ze^- \rightarrow M_{(s)}$$

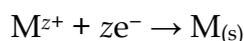

using the difference between metal SA adsorption energy and the cohesive energy of the corresponding bulk metal. Used standard electrode potentials for the $M^{z+}/M$ couples are also provided in reference [24]. Note that the data for the $Ir^{z+}/Ir$ couple are not available. The results for the cases of Ni, Pd and Pt SA are presented in Figure 7. Calculated standard electrode potentials for all the $E_{Mz+/M@support}$ couples (except the Ir case) are provided in Supplementary information, Table S1.

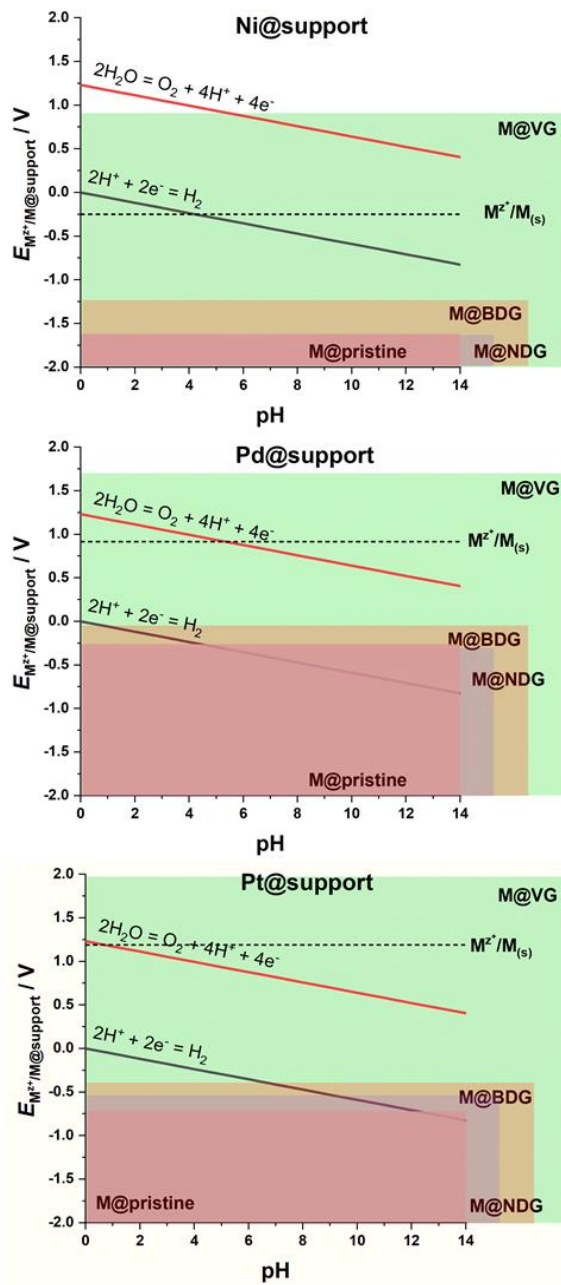

**Figure 7.** Stability of Ni, Pd and Pt in terms of dissolution from the support in the form of M$^{z+}$ ions. In all three cases $z = 2$.

The obtained results suggest that, in general, metal SA supported by pristine graphene, NDG, and BDG are more prone to dissolution than the corresponding clean bulk metals. Even noble metals like Ag and Au dissolve at potentials that do not cover the entire electrochemical window at every pH. However, when metal SAs are embedded in graphene at the vacancy site, they become more stable than bulk metals, except Ag and Au, which interact rather weakly with the single vacancy site in graphene (Table S1). While Ni and Cu are stable at high pH, Ru, Rh, Pd and Pt should never dissolve from the vacancy site. Hence, these metal SA embedded in the vacancy should be stable at both SAC|vacuum interface and SAC|aqueous electrolyte interface (in the pH range 0 to 14). In terms of stability, these four model SACs seem to have everything required for the application in practice – they should be easy to make (aggregation is not preferred), while metal corrosion is suppressed.

**3.2. Adsorption trends for atoms and molecules on M@support model SACs**

Proper interactions of a catalyst with reactants and reaction intermediates are crucial for efficient catalysis. Hence, understand adsorption and reactivity trends could provide valuable insights into the design of new catalytic materials. When it comes to solid metallic surfaces, adsorption and reactivity (and consequently catalytic activity) trends were efficiently linked to the electronic structure. However, the analysis of DOS (Figure 3 and 4, and Figure S1 and S2) suggests that the $d$-states of metal SAs supported on pristine graphene, NDG, and BDG are highly localized, resembling an atomic or molecular system, while in the case of VG metal $d$-states are wide and metal-like. If the reactivity is linked to the electronic structure, one would expect quite different reactivity of a given metal atom deposited on VG, on one side, and other supports investigated here, on the other side.

We have analyzed atomic adsorption of H, C, N, O and S, and $AH_x$ molecules (A = C, N, O, or S; 11 different species with various levels of A saturation by H) at model SACs for all nine metals and four different supports, resulting with the total of 576 different systems. We note that these atoms and molecules are chosen due to their importance in catalysis. H and $OH_x$ are relevant for the hydrogen and oxygen electrode reactions, N and $NH_x$ due to synthesis $NH_3$ and electrochemical reduction of $N_2$, $CH_x$ species are intermediates in numerous catalytic and electrocatalytic reactions, while sulfur species are known as catalytic poisons. Considering the potential importance of M@VG systems, demonstrated in Section 3.1, here we report the data obtained for this family of SACs, while the data for other considered supports are provided in the Supplementary Information (Tables S2-S4).

**Table 2.** Adsorption energies of H, A (A = C, N, O and S) on M@VG model SACs.

| M  | H     | X     | XH    | XH$_2$ | XH$_3$ | XH$_4$ |
|----|-------|-------|-------|--------|--------|--------|
|    |       |       |   C   |        |        |        |
| Ni | −2.01 | −3.34 | −3.560 | −3.62 | −1.89  | −0.10  |
| Cu | −2.20 | −3.29 | −3.83 | −3.65  | −2.00  | −0.13  |
| Ru | −2.44 | −3.58 | −3.84 | −3.72  | −2.34  | −0.24  |
| Rh | −2.65 | −3.30 | −3.68 | −3.85  | −2.52  | −0.15  |
| Pd | −2.05 | −3.04 | −3.27 | −3.35  | −1.64  | −0.10  |
| Ag | −2.26 | −2.51 | −3.11 | −3.26  | −1.57  | −0.15  |
| Ir | −3.26 | −3.87 | −4.44 | −4.57  | −3.15  | −0.11  |
| Pt | −2.44 | −3.72 | −4.13 | −4.19  | −2.26  | −0.11  |
| Au | −3.25 | −4.34 | −4.93 | −4.56  | −2.76  | −0.27  |
|    |       |       |   N   |        |        |        |
| Ni |       | −2.50 | −3.00 | −2.70  | −0.73  |        |
| Cu |       | −2.42 | −2.79 | −2.68  | −0.96  |        |
| Ru |       | −2.36 | −2.71 | −2.85  | −1.34  |        |
| Rh |       | −2.24 | −2.63 | −3.00  | −0.76  |        |
| Pd |       | −1.94 | −2.51 | −2.35  | −0.66  |        |
| Ag |       | −1.26 | −1.88 | −2.49  | −0.92  |        |
| Ir |       | −2.99 | −3.37 | −3.64  | −0.72  |        |
| Pt |       | −2.73 | −3.36 | −2.79  | −0.46  |        |
| Au |       | −3.72 | −3.70 | −3.52  | −1.52  |        |
|    |       |       |   O   |        |        |        |
| Ni |       | −4.46 | −3.75 | −0.36  |        |        |
| Cu |       | −3.84 | −3.43 | −0.52  |        |        |
| Ru |       | −3.85 | −3.83 | −0.79  |        |        |
| Rh |       | −3.97 | −3.91 | −0.40  |        |        |
| Pd |       | −3.92 | −3.34 | −0.32  |        |        |
| Ag |       | −2.97 | −3.28 | −0.50  |        |        |
| Ir |       | −4.90 | −4.49 | −0.27  |        |        |
| Pt |       | −4.75 | −3.65 | −0.19  |        |        |
| Au |       | −4.89 | −4.14 | −0.48  |        |        |
|    |       |       |   S   |        |        |        |
| Ni |       | −3.78 | −2.91 | −0.62  |        |        |
| Cu |       | −3.46 | −3.01 | −0.73  |        |        |
| Ru |       | −3.27 | −3.09 | −1.09  |        |        |
| Rh |       | −3.34 | −3.30 | −0.76  |        |        |
| Pd |       | −3.46 | −2.81 | −0.45  |        |        |
| Ag |       | −3.00 | −2.87 | −0.74  |        |        |
| Ir |       | −4.10 | −3.84 | −0.83  |        |        |
| Pt |       | −4.28 | −3.21 | −0.72  |        |        |
| Au |       | −4.62 | −3.74 | −1.12  |        |        |

The results show a strong interaction between metal SA on all of the studied supports and the investigated adsorbates that are not fully saturated by hydrogen. It is important to observe that the $E_{ads}$(A) on metal SAs do not correlate to the corresponding adsorption energies of the same adsorbates on the bulk metallic surfaces found in the literature.[34] For example, in the group of M@VG, Au SAs are the most reactive towards atomic adsorbates. However, when on the other supports, Au SAs do not interact so strongly with studied adsorbates. Another important difference between supported SAs and the corresponding bulk metallic surfaces is that is no general weakening of the interaction between SAs and adsorbates as they get more saturated by hydrogen, but certain trends are definitely visible.[34] As expected, the interactions of model SAs with fully saturated molecular adsorbates ($CH_4$, $NH_3$, $OH_2$, and $SH_2$) are rather weak.

Figure 8 shows the scaling between $E_{ads}(A)$ and $E_{ads}(AH_x)$ for metal SAs at four different supports (except for C $vs$. $CH_3$). As can be seen from Figure S7-S10, Supplementary Information, where scaling relations are shown separately for different supports, the scaling between $E_{ads}(A)$, on one side and $E_{ads}(AH)$ or $E_{ads}(AH_2)$, on the other side, is quite good, particularly on M@VG. However, for $CH_3$ and fully saturated $AH_x$, the scaling relations are broken. For all the adsorbates, except the broken scaling cases, the statistics says that the correlation exists (at the 0.05 level). We must note that it is not surprising that scaling does not work for the fully saturated $AH_x$, as the bonding to the M sites is essentially different than in the case of atoms and molecular fragment (see Section 3.4). The theory of Abil-Pedersen $et\ al.$[34] obviously cannot be applied here directly, which we consider a consequence of a partial atom- or molecule-like character of metal SAs on studied supports, which can be concluded from presented DOS plots (Figures 3 and 4). However, it is important to mention that previously scaling relations for O-containing species relevant for oxygen reduction reaction have not been found on metal SAs embedded in graphene lattice.[35] In contrast, new insights on the reactivity of graphene with heteroatoms show that such relations could exist.[36] The parameters of scaling relations presented in Figure 8 are given in Table S5. Obviously, the slopes of the scaling relations group around 0.4 and 0.7. This observation needs further theoretical treatment to provide justification.

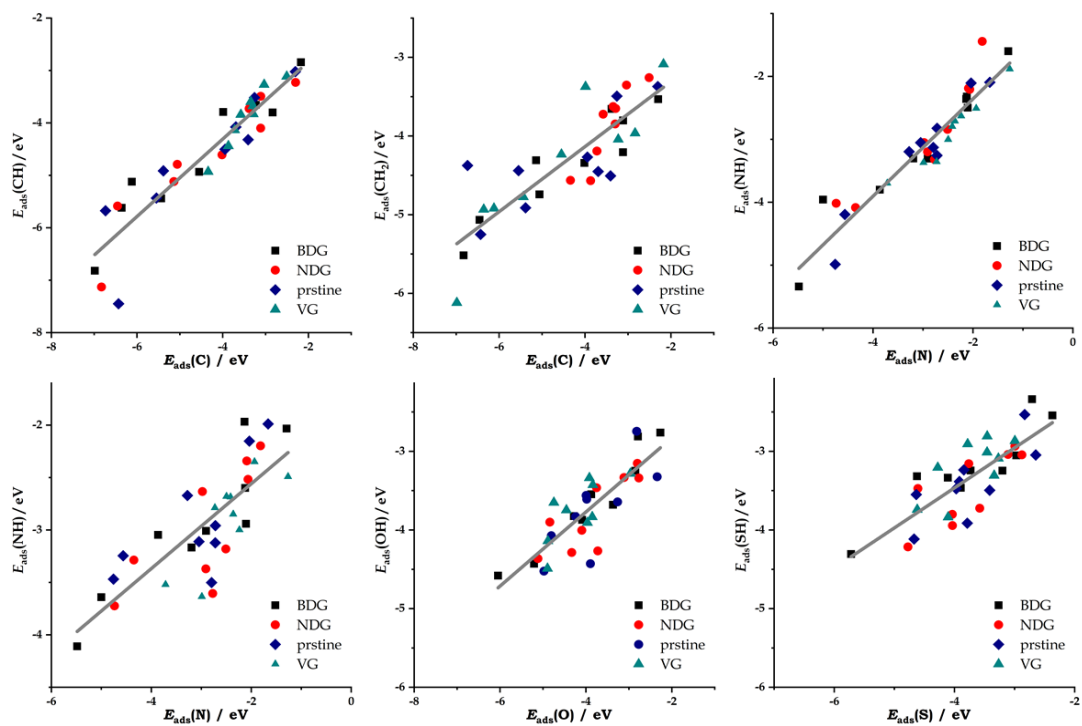

**Figure 8.** Scaling relations between $E_{ads}(A)$ and $E_{ads}(AH_x)$ ($x$ = 1, 2 for A = C, N, $x$ = 1 for A = O, S) for metal SAs on four studied supports. Gray lines are linear fits of presented data. Parameters of the fit are provided in Table S5, Supplementary information.

To confirm a general connection between the electronic structure and the reactivity of metal SAs on different support, another type of scaling can be checked. Namely, metal SAs interact relatively weakly with pristine graphene, as well as NDG and BDG (Table 1). As a result, DOSes of SAs on these supports are very similar (Figure 3 and Figure S1-S4, Supplementary information). Hence, the similarity of the electronic structure should result in similar reactivity. This is indeed confirmed when adsorption energies on M@pristine graphene are correlated with the adsorption energies on M@NDG and M@BDG (Figure 9, top and middle left). Much stronger interactions of metal SAs with the vacancy site (Table 1) results in a more disrupted electronic structure of SAs, and the correlation between adsorption energies of atomic and molecular adsorbates on M@VG scales much poorer with the corresponding adsorption energies on M@pristine graphene (Figure 9, bottom left). First, one should look at the integrated local density of states (ILDOS) of Pt and Au SAs at different supports (Figure 9, right; see also Figure S11 that shows that the integration is done in the energy window where SAs *d*-states are located). It becomes clear that Au and Pt (and other SAs) on pristine graphene, NDG and BDG are atom-like as the interaction with the supports relatively weakly disrupts the electronic structure of SAs. However, the d-states of SAs on VG are strongly hybridized with the graphene π system and metal-like, as indicated by Figure 3. Hence, as a rule of thumb, one can say that M@NDG and M@BDG bind studied adsorbates like M@pristine graphene. The dopant in graphene lattice gently modifies metal anchoring on the support.

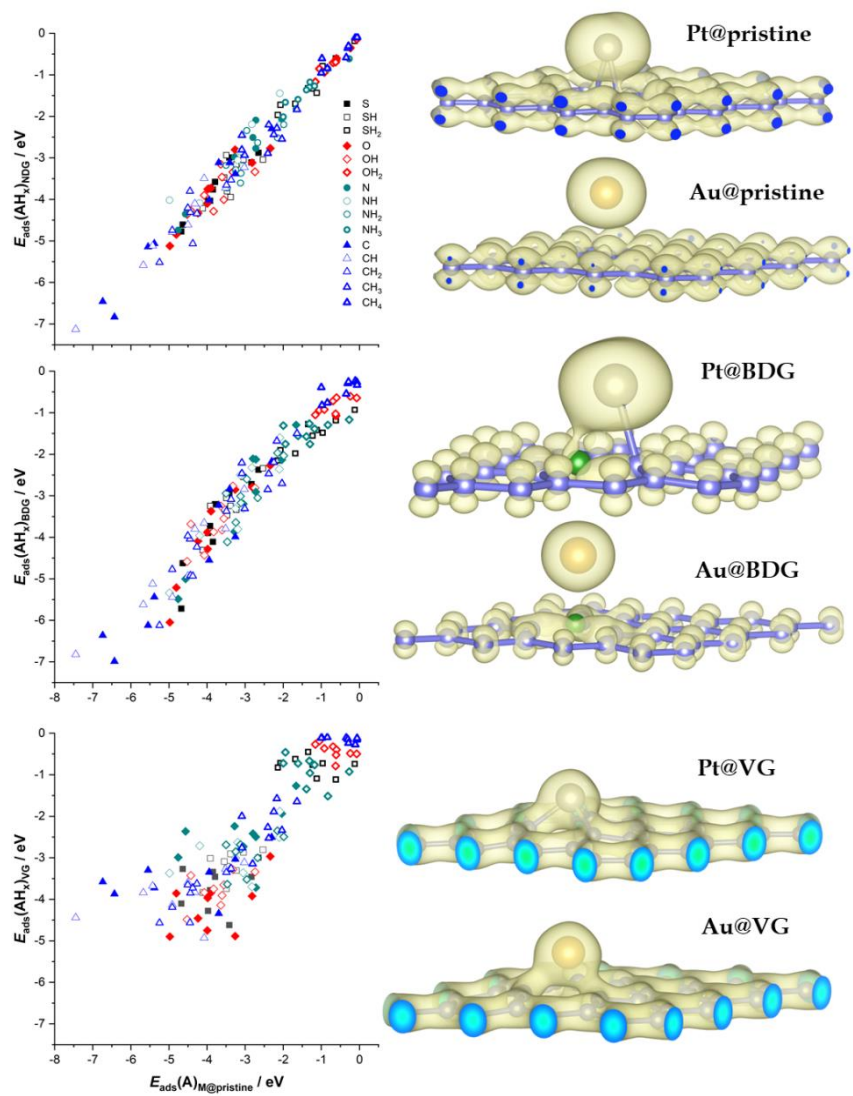

**Figure 9.** Correlation of atomic and molecular adsorption energies on M@pristine graphene ad M@NDG (top left), M@BDG (middle left) and M@VG (bottom left). On the right, isosurfaces of integrated local densities of states (ILDOS) are presented for selected supported metals SAs. Integration was performed in the energy windows that cover the metal the *d*-states (see Figure S11, and Figures 3 and 4).

### 3.3. Catalysis on model SACs – the case of hydrogen evolution reaction (HER)

Provided datasets of adsorption energies of investigated simple adsorbates can be used to analyze different chemical and electrochemical processes on the model SACs. Here we look into more details a relatively simple case of electrocatalytic HER on studied model SACs. Current views suggest that, in a simple model, it is possible to relate the catalytic activity to the energetics of hydrogen adsorption on the catalytically active site, and we follow that approach.[37] Considering that the thermodynamic stability of model SACs was already investigated (Section 3.1) and that only M@VG systems are not prone to the dissolution in highly acidic solutions (except Ag@VG and Au@VG), we focus on this family of SACs. We converted $E_{ads}$(H) to the free energy of the formation of adsorbed H ($\Delta G_{ads}$(H)), which is the intermediate for HER, in order to obtain HER reaction energy profiles on M@VG (Figure 10, left). The reactivity of SAs is altered in comparison to the parental extended metal surfaces. Hence, the trends in HER activity, depicted *via* the volcano curve (Figure 10, right), are also rather different. For example, Pt, Ir and Pd, take their place close to the apex of HER volcano in the case of the extended metal surfaces. However, in the case of M@VG SACs, only Pt shows high activity. Ir is found at the ascending branch due to the very strong bonding of $H_{ads}$, while Pd is at the descending branch as it binds $H_{ads}$ weakly. Hence, in the case of Pd (and other metals on the descending branch), the Volmer reaction, i.e., the formation of $H_{ads}$ [38]:

M@VG + H$^+$ + e$^-$ → $H_{ads}$-M@VG  [Mechanism 1]

is the potential determining step for HER. For the metals at the ascending branch, $H_{ads}$ removal controls HER. In the case of extended metal surfaces, $H_{ads}$ can be removed by either Tafel reaction:

$H_{ads}$ + $H_{ads}$ → $H_2$↑  [Mechanism 2]

or Heyrovsky reaction [38]:

$H_{ads}$ + H$^+$ + e$^-$ → $H_2$↑  [Mechanism 3]

As the reaction [Mechanism 2] requires two $H_{ads}$ adjacent to each other to recombine, $H_2$ formation should go through the reaction [Mechanism 3], irrespective of the position of SAC on the HER volcano. This conclusion holds if a metal SA site binds only one H during HER, which is likely for the metals at the descending branch.

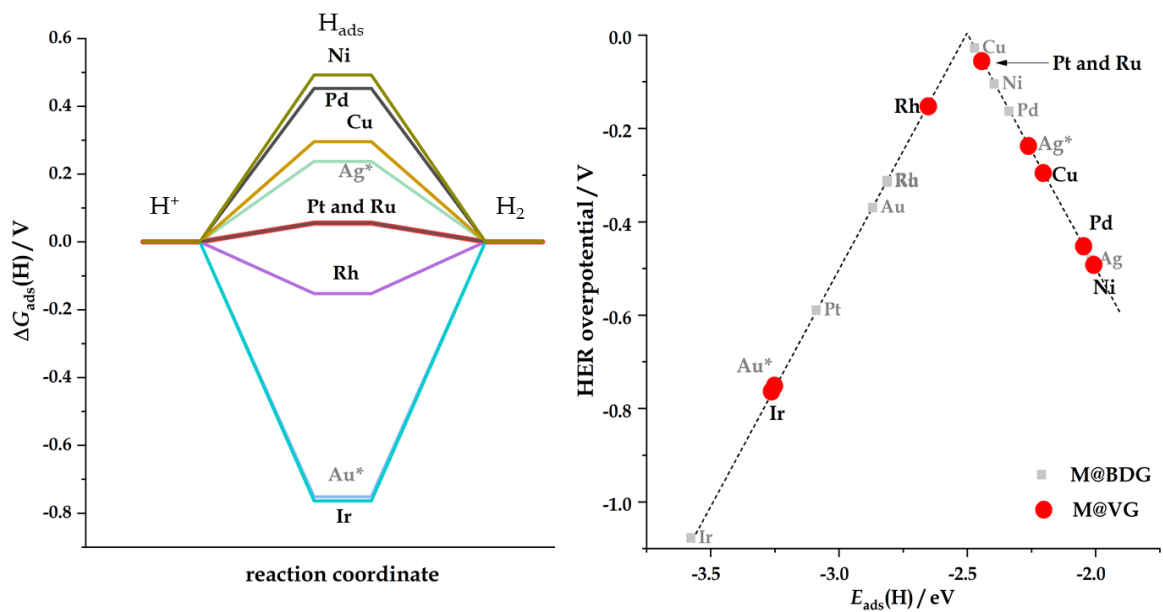

**Figure 10.** HER catalytic activity of studied model SACs – Reaction free energy profile for HER on M@VG systems (left) and HER volcano plot that is assembling M@VG (circles) and M@BDG (squares). For the M@VG family, metal SAs which dissolve at pH = 0 are indicated by (*). All of the M@BGD dissolve at pH = 0.

The results indicate that Pt@VG and Ru@VG have the highest HER activities. This result is not exciting as both Pt and Ru are scarce and expensive, but using SACs could still be beneficial due to the lowering of the metal loading in the catalyst. In this sense, M@BDG is more attractive as Cu@BDG should show even better HER activity than Pt@VG and Ru@VG (Figure 10, right), but all the M@BDG dissolve at pH = 0.

One must not forget that practically each metal SA is at the interface with the support. Hence, additional interfacial effects can boost $H_2$ production. This particularly relates to $H_{ads}$ spillover[39,40], which is beneficial when strong interactions with active metal sites exist.[41] We have previously shown that defects in graphene lattice (and metals SAs are, in this sense, defects) usually increase the reactivity of the graphene basal plane in the vicinity of a defect.[42-44] Hence, $H_{ads}$ could spill to the graphene plane around the defect ($C_x$):

$H_{ads}$-M@VG + $C_x$ → M@VG + $H_{ads}$-$C_x$         [Mechanism 4]

and H$_2$ could form via recombination from the support around the defect:

H$_{ads}$-C$_x$ + H$_{ads}$-C$_x$ → H$_2$↑ + C$_x$ + C$_x$     [Mechanism 5]

While we do not address this possibility here, it is important to note that the H$_{ads}$ spillover could be greatly beneficial for HER as it provided cleaning action for metal SACs. Nevertheless, it should be generally considered as in SACs; all the catalytic sites are interfacial ones. Besides, due to the altered reactivity of the support in the vicinity of metal SA, the energetics of H$_{ads}$ formation could be altered so that the support itself becomes catalytically active so that reactions [Mechanism 1]-[Mechanism 3] could proceed directly on the support. All these effects are yet to be analyzed in detail. However, we briefly investigated a few systems of the M@VG family, which are at the apex of the Volcano curve in Fig. 10 (right). For the cases of Pt@VG and Ru@VG, the H adsorption energies at the C sites next to the metal SAC are −1.57 eV and −1.63 eV, respectively. Hence, the formation of H$_{ads}$ is energetically unfavorable, and H$_{ads}$ spillover from metal SAC to the support is unlikely to occur. However, in the case of Cu@VG, $E_{ads}$(H) at the C atom of the vacancy site amounts to −2.52 eV. This value is almost perfect $E_{ads}$(H) regarding maximizing HER activity. Hence, if HER is also considered at the C atoms of the vacancy, Cu@VG would be the most active catalyst among considered ones. Moreover, H$_{ads}$ formation at the Cu site is followed by −2.20 eV (Table 2), so H$_{ads}$ can transfer from the support to Cu atom in the reverse spillover process analogous to [Mechanisms 4-5]. The thermodynamic barrier for this process is roughly 0.32 eV, which can be overcome at room temperature. As we observed that in the case of Cu@VG, H binds stronger to the C site in the vacancy, we checked the same for all considered metals. Such behavior was also seen for Ag@VG and Au@VG (Table S6). Without any doubts, this is ascribed to the weak bonding of Cu, Ag and Au at the vacancy site (Table 1), so the C atoms are not as saturated as in other cases. In other cases, H$_{ads}$ forms at the metal sites. Finally, in the case of Rh@VG, the H adsorption energy at the C site of the vacancy is −2.04 eV. Hence, H$_{ads}$ might migrate from Rh single atom to the C sites in the vicinity with a thermodynamic barrier of 0.61 eV. This barrier is relatively large, but it can be overcome at room temperature or higher temperatures. With the results above, the apex of the volcano curve (Fig. 10) can be redrawn, as presented in

**Figure 11**. Different possibilities where HER take place in parallel are indicated, like in Ref. [45]

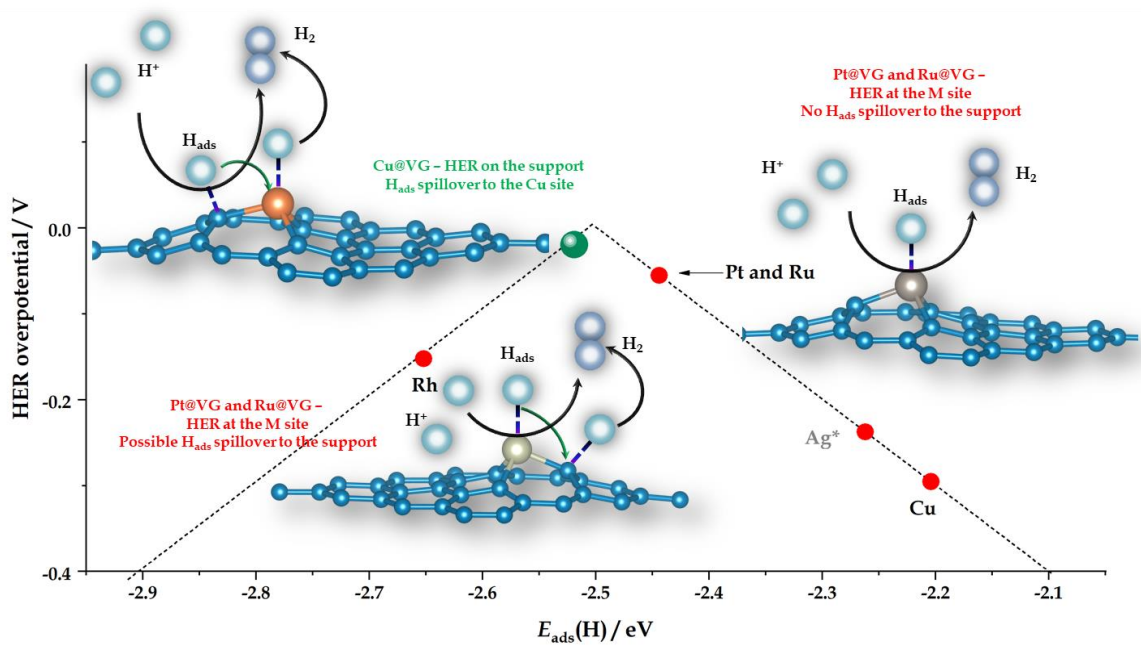

**Figure 11.** The apex of the HER volcano, where different reaction pathways are presented. In the case of Rh@VG, HER takes place at the metal site, while Hads spillover on the support is possible. In the case of Cu@VG, active sites for HER are the C sites around the vacancy, while Hads can spill onto the Cu site. For Pt@VG and Ru@VG HER, the active sites for HER are only metal sites. The positions of the SACs are determined based on the $E_{ads}$(H) at the most stable site.

### 3.4. Additional computational aspects

As graphene is generally chemically inert, taking dispersion interactions into account when theoretically treated is very important.[30,46] In this contribution, we used the PBE-D2 approach consistently, which is not of the highest importance when strong chemical bonds are present (M-A and M-AH$_x$ interaction), but it is taken into account because of the graphene-metal interaction (particular cases of pristine graphene, NDG and BDG). In this sense, the importance of dispersion interactions is visible in Table 1, where $E_{ads}$(M) are reported. The contributions are up to 0.6 eV, which in some cases presents a large portion of $E_{ads}$(M). When adsorption of different atomic and molecular species at the active sites of SACs is considered, the effects are also visible and amount up to 0.3 eV (Figure S12). This value is somewhat smaller than the absolute impact on the $E_{ads}$(M), but relative contributions of dispersion interactions are extremely high in the case of fully saturated adsorbates ($CH_4$, $NH_3$, $OH_2$ and $SH_2$), which is not surprising. The largest contribution is in the case of $CH_4$ because dispersion interactions are practically the only ones that account in this case. $NH_3$, $OH_2$ and $SH_2$ have lone electron pairs and are polar molecules so that the interactions can be electrostatic, or the lone electron pairs can interact directly with the *d*-states of metal SAs. Considering the results above, we conclude that dispersion should always be considered when graphene or modified graphene surfaces are an integral part of the system under the investigation.

## 4. Conclusions

We considered nine metals frequently used in catalysis and electrocatalysis (Ni, Cu, Ru, Rh, Pd, Ag, Ir, Pt and Au) and their employment in SACs with graphene-based supports. Substitutional doping of pristine graphene with B and N modified slightly modifies metal SAs anchoring to chemically inert graphene basal plane. On the other side, a single vacancy presents the site of very strong bonding. As a result, stable SACs (except in the case of Ag and Au), which are not expected to aggregate or dissolve under harsh electrochemical conditions, can be obtained. Using atomic (hydrogen and A = C, N, O and S) and molecular adsorbates ($AH_x$, $x$ = 1, 2, 3 or 4, depending on A), we screened the reactivity of model SACs, resulting in scaling relations (except in the case of $CH_3$). Moreover, the similarity of the electronic structures of metals SAs on pristine, N-doped and B-doped graphene results in clear scaling between adsorption energies of atomic and molecular adsorbates on metals SAs supported by these surfaces. These conclusions can be used to rapidly screen considered SACs for different catalytic processes that involve studied atomic and molecular species. Using the obtained data for hydrogen adsorption energetics on metal SAs considered model SACs were analyzed in terms of HER activity. The only promising systems, in this case, are M@VG, and Pt@VG and Ru@VG stand out as the most active SACs for HER. However, if the contribution of the support is also considered, the most active catalyst is Cu@VG.

In real life, nanosized particles are very difficult for theoretical treatment as even very small nanoparticles have a relatively large number of atoms compared to the capabilities of electronic structure calculations. Hence, some simplifications must be taken into account. In this sense, relative simplicity and the sizes of model SACs make them a dream comes true for theoretical analysis. However, additional effects must be considered when analyzing SACs, which are less critical for nanosized catalysts. These are primarily interfacial effects, as every SA is the interfacial atom. Both contributions of the support and altered reactivity of the support are essential to completely understand and explain or predict the catalytic performance of SACs. Moreover, with (modified) graphene as the SA support and molecular adsorbates which lack dangling bonds of lone electron couples that allow strong chemical bonding to SAs, dispersion interactions should be considered when analyzing (re)activity of model systems using DFT.

**Acknowledgment**

This research was supported by the Serbian Ministry of Education, Science and Technological Development (Contract number: 451-03-68/2020-14/200146) and the Science Fund of the Republic of Serbia (PROMIS programme, RatioCAT, no. 606224). This research was sponsored in part by the NATO Science for Peace and Security Programme under grant no. G5729. S.V.M acknowledges the support provided by the Serbian Academy of Sciences and Arts for funding the study through the project "Electrocatalysis in the contemporary processes of energy conversion". N.V.S. acknowledges the support provided by the Swedish Research Council through the project no. 2014-5993. The computations were performed on resources provided by the Swedish National Infrastructure for Computing (SNIC) at National Supercomputer Centre (NSC) at Linköping University.

**References**

[1]    M. S. Dresselhaus, I. L. Thomas, *Nature* **2001**, DOI 10.1038/35104599.

[2]    K. Sasaki, H. Naohara, Y. Cai, Y. M. Choi, P. Liu, M. B. Vukmirovic, J. X.

Wang, R. R. Adzic, *Angew. Chemie - Int. Ed.* **2010**, DOI 10.1002/anie.201004287.

[3]   J. Zhang, F. H. B. Lima, M. H. Shao, K. Sasaki, J. X. Wang, J. Hanson, R. R. Adzic, *J. Phys. Chem. B* **2005**, DOI 10.1021/jp055634c.

[4]   X. F. Yang, A. Wang, B. Qiao, J. Li, J. Liu, T. Zhang, *Acc. Chem. Res.* **2013**, DOI 10.1021/ar300361m.

[5]   B. Qiao, A. Wang, X. Yang, L. F. Allard, Z. Jiang, Y. Cui, J. Liu, J. Li, T. Zhang, *Nat. Chem.* **2011**, DOI 10.1038/nchem.1095.

[6]   Y. Chen, S. Ji, Y. Wang, J. Dong, W. Chen, Z. Li, R. Shen, L. Zheng, Z. Zhuang, D. Wang, Y. Li, *Angew. Chemie - Int. Ed.* **2017**, DOI 10.1002/anie.201702473.

[7]   H. T. Chung, D. A. Cullen, D. Higgins, B. T. Sneed, E. F. Holby, K. L. More, P. Zelenay, *Science (80-. ).* **2017**, DOI 10.1126/science.aan2255.

[8]   F. Chen, X. Jiang, L. Zhang, R. Lang, B. Qiao, *Cuihua Xuebao/Chinese J. Catal.* **2018**, DOI 10.1016/S1872-2067(18)63047-5.

[9]   H. Zhang, G. Liu, L. Shi, J. Ye, *Adv. Energy Mater.* **2018**, DOI 10.1002/aenm.201701343.

[10]  A. Wang, J. Li, T. Zhang, *Nat. Rev. Chem.* **2018**, DOI 10.1038/s41570-018-0010-1.

[11]  R. T. Hannagan, G. Giannakakis, M. Flytzani-Stephanopoulos, E. C. H. Sykes, *Chem. Rev.* **2020**, DOI 10.1021/acs.chemrev.0c00078.

[12]  N. Cheng, L. Zhang, K. Doyle-Davis, X. Sun, *Electrochem. Energy Rev.* **2019**, DOI 10.1007/s41918-019-00050-6.

[13]  H. Yan, C. Su, J. He, W. Chen, *J. Mater. Chem. A* **2018**, DOI 10.1039/c8ta01940a.

[14]  M. Li, H. Wang, W. Luo, P. C. Sherrell, J. Chen, J. Yang, *Adv. Mater.* **2020**, DOI 10.1002/adma.202001848.

[15]  H. Liu, X. Peng, X. Liu, *ChemElectroChem* **2018**, DOI 10.1002/celc.201800507.

[16]  Y. Chen, S. Ji, C. Chen, Q. Peng, D. Wang, Y. Li, *Joule* **2018**, DOI 10.1016/j.joule.2018.06.019.

[17]  L. Zhang, Y. Ren, W. Liu, A. Wang, T. Zhang, *Natl. Sci. Rev.* **2018**, DOI 10.1093/nsr/nwy077.

[18]  C. Rivera-Cárcamo, P. Serp, *ChemCatChem* **2018**, DOI 10.1002/cctc.201801174.

[19]  H. Xu, D. Cheng, D. Cao, X. C. Zeng, *Nat. Catal.* **2018**, DOI 10.1038/s41929-018-0063-z.

[20] J. P. Perdew, K. Burke, M. Ernzerhof, D. of Physics, N. O. L. 70118 J. Quantum Theory Group Tulane University, *Phys. Rev. Lett.* **1996**, *77*, 3865.

[21] P. Giannozzi, S. Baroni, N. Bonini, M. Calandra, R. Car, C. Cavazzoni, D. Ceresoli, G. L. Chiarotti, M. Cococcioni, I. Dabo, A. Dal Corso, S. de Gironcoli, S. Fabris, G. Fratesi, R. Gebauer, U. Gerstmann, C. Gougoussis, A. Kokalj, M. Lazzeri, L. Martin-Samos, N. Marzari, F. Mauri, R. Mazzarello, S. Paolini, A. Pasquarello, L. Paulatto, C. Sbraccia, S. Scandolo, G. Sclauzero, A. P. Seitsonen, A. Smogunov, P. Umari, R. M. Wentzcovitch, *J. Phys. Condens. Matter* **2009**, *21*, 395502.

[22] S. Grimme, *J. Comput. Chem.* **2006**, *27*, 1787.

[23] H. J. Monkhorst, J. D. Pack, *Phys. Rev. B* **1976**, *13*, 5188.

[24] B. Johansson, A. Jovanović, S. V. Mentus, I. A. Pašti, N. V. Skorodumova, A. S. Dobrota, *Phys. Chem. Chem. Phys.* **2017**, *20*, 858.

[25] H. Shin, S. Kang, J. Koo, H. Lee, J. Kim, Y. Kwon, *J. Chem. Phys.* **2014**, *140*, 114702.

[26] R. W. Lynch, H. G. Drickamer, *J. Chem. Phys.* **1966**, *44*, 181.

[27] S. T. Skowron, I. V. Lebedeva, A. M. Popov, E. Bichoutskaia, *Chem. Soc. Rev.* **2015**, *44*, 3143.

[28] A. El-Barbary, H. Telling, P. Ewels, I. Heggie, R. Briddon, *Phys. Rev. B - Condens. Matter Mater. Phys.* **2003**, *68*, 144107.

[29] L. Rodrigo, P. Pou, R. Pérez, *Carbon N. Y.* **2016**, *103*, 200.

[30] A. Jovanović, S. V. Mentus, B. Johansson, A. S. Dobrota, N. V. Skorodumova, I. A. Pašti, *Appl. Surf. Sci.* **2017**, *436*, 433.

[31] K. T. Chan, J. B. Neaton, M. L. Cohen, *Phys. Rev. B - Condens. Matter Mater. Phys.* **2008**, DOI 10.1103/PhysRevB.77.235430.

[32] M. Amft, S. Lebègue, O. Eriksson, N. V. Skorodumova, *J. Phys. Condens. Matter* **2011**, DOI 10.1088/0953-8984/23/39/395001.

[33] C. Kittel, *Am. J. Phys.* **1957**, *25*, 330.

[34] F. Abild-Pedersen, J. Greeley, F. Studt, J. Rossmeisl, T. R. Munter, P. G. Moses, E. Skúlason, T. Bligaard, J. K. Nørskov, *Phys. Rev. Lett.* **2007**, *99*, 016105.

[35] M. Kaukonen, A. V. Krasheninnikov, E. Kauppinen, R. M. Nieminen, *ACS*

*Catal.* **2013**, *3*, 159.

[36] S. V. Doronin, A. A. Volykhov, A. I. Inozemtseva, D. Y. Usachov, L. V. Yashina, *J. Phys. Chem. C* **2020**, DOI 10.1021/acs.jpcc.9b09668.

[37] J. K. Nørskov, T. Bligaard, A. Logadottir, J. R. Kitchin, J. G. Chen, S. Pandelov, U. Stimming, *J. Electrochem. Soc.* **2005**, *152*, J23.

[38] J. O. M. Bockris, D. F. A. Koch, *J. Phys. Chem.* **1961**, DOI 10.1021/j100828a007.

[39] R. Prins, *Chem. Rev.* **2012**, *112*, 2714.

[40] W. C. Conner, J. L. Falconer, *Chem. Rev.* **1995**, *95*, 759.

[41] I. A. Pašti, M. Leetmaa, N. V Skorodumova, *Int. J. Hydrogen Energy* **2016**, *41*, 2526.

[42] A. S. Dobrota, I. A. Pašti, S. V. Mentus, B. Johansson, N. V. Skorodumova, *Appl. Surf. Sci.* **2020**, *514*, 145937.

[43] N. P. Diklić, A. S. Dobrota, I. A. Pašti, S. V. Mentus, B. Johansson, N. V. Skorodumova, *Electrochim. Acta* **2019**, *297*, 523.

[44] A. S. Dobrota, I. A. Pašti, S. V. Mentus, N. V. Skorodumova, *Phys. Chem. Chem. Phys.* **2017**, *19*, 8530.

[45] G. Vilé, D. Albani, M. Nachtegaal, Z. Chen, D. Dontsova, M. Antonietti, N. López, J. Pérez-Ramírez, *Angew. Chemie - Int. Ed.* **2015**, DOI 10.1002/anie.201505073.

[46] K. A. Novčić, A. S. Dobrota, M. Petković, B. Johansson, N. V. Skorodumova, S. V. Mentus, I. A. Pašti, *Electrochim. Acta* **2020**, *354*, 136735.

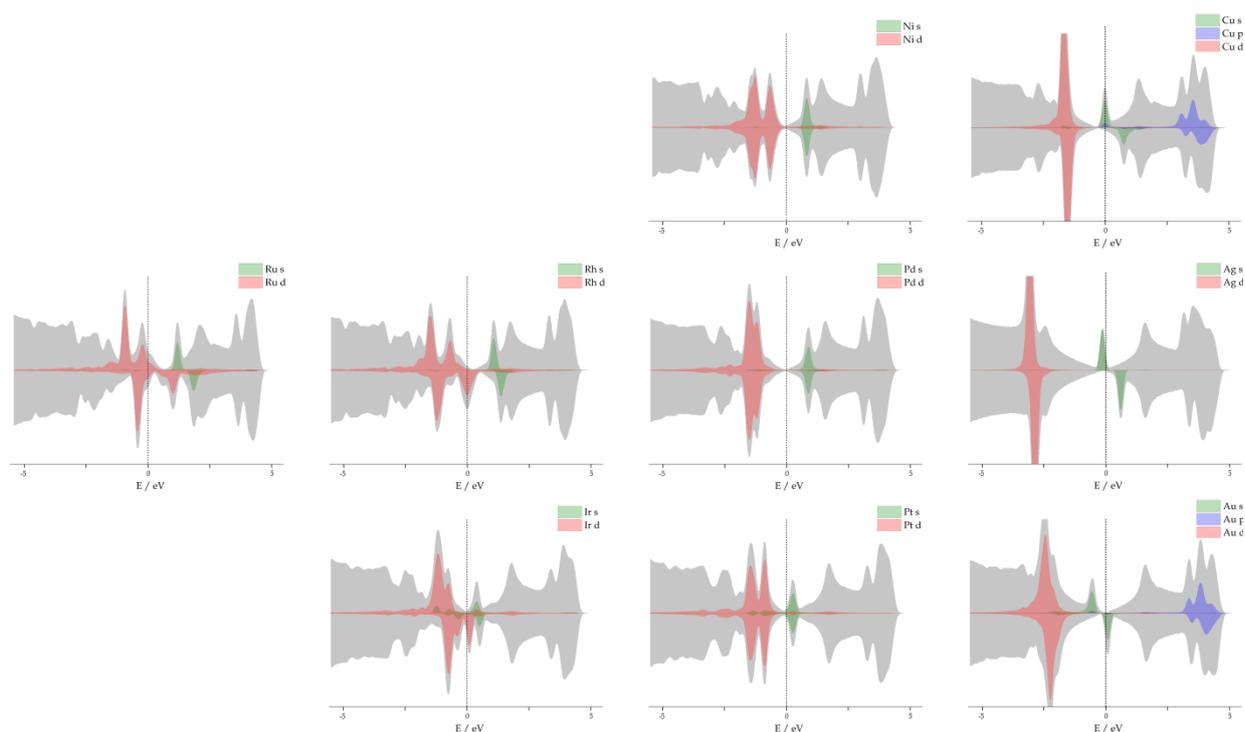

**Figure S1.** DOS plots for single atoms adsorbed on pristine graphene. The energy scale is referred to the Fermi level (vertical dashed line).

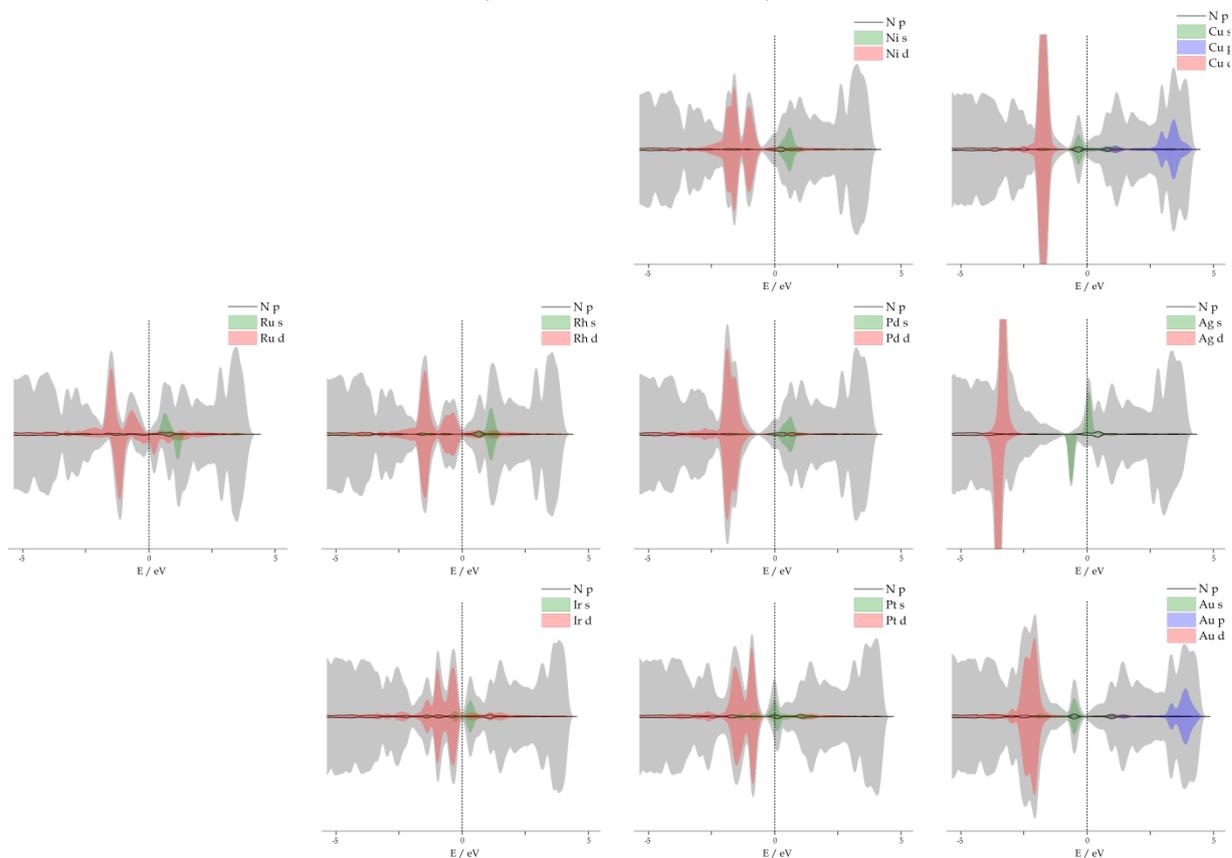

**Figure S2.** DOS plots for single atoms adsorbed on nitrogen-doped graphene. The energy scale is referred to the Fermi level (vertical dashed line).

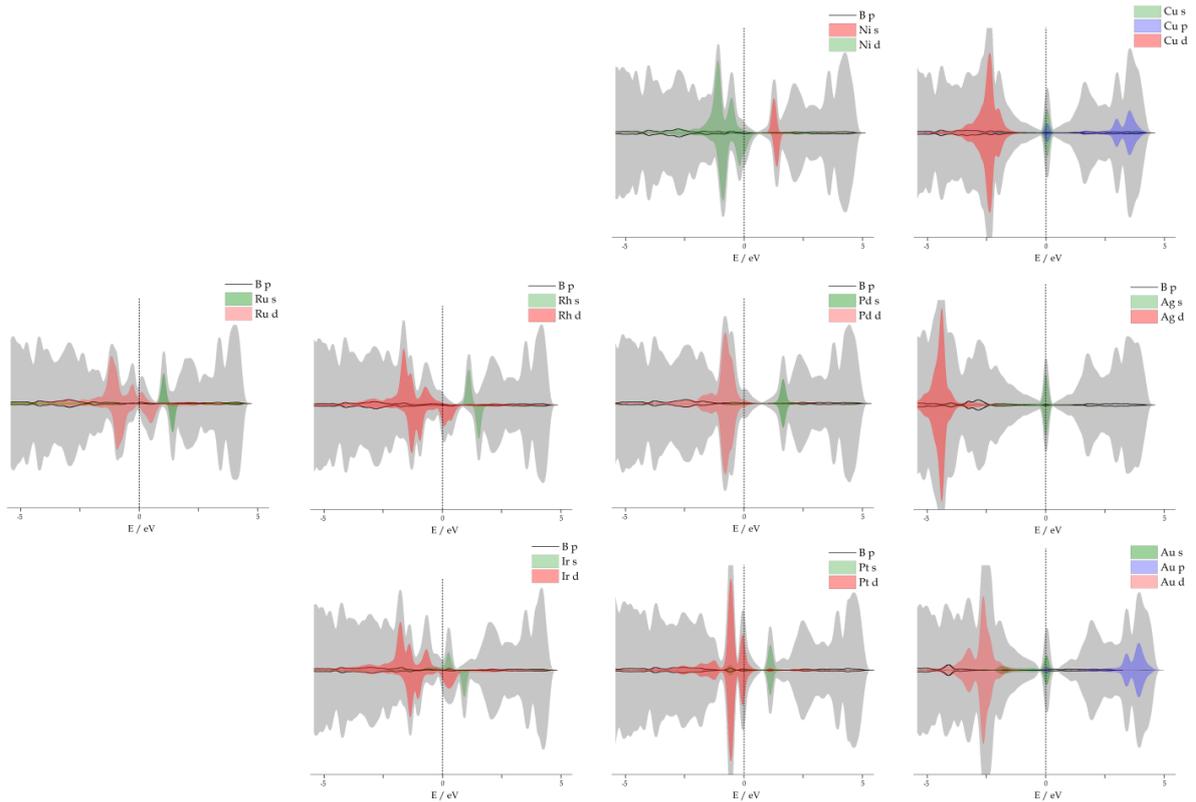

**Figure S3.** DOS plots for single atoms adsorbed on boron−doped graphene. The energy scale is referred to the Fermi level (vertical dashed line).

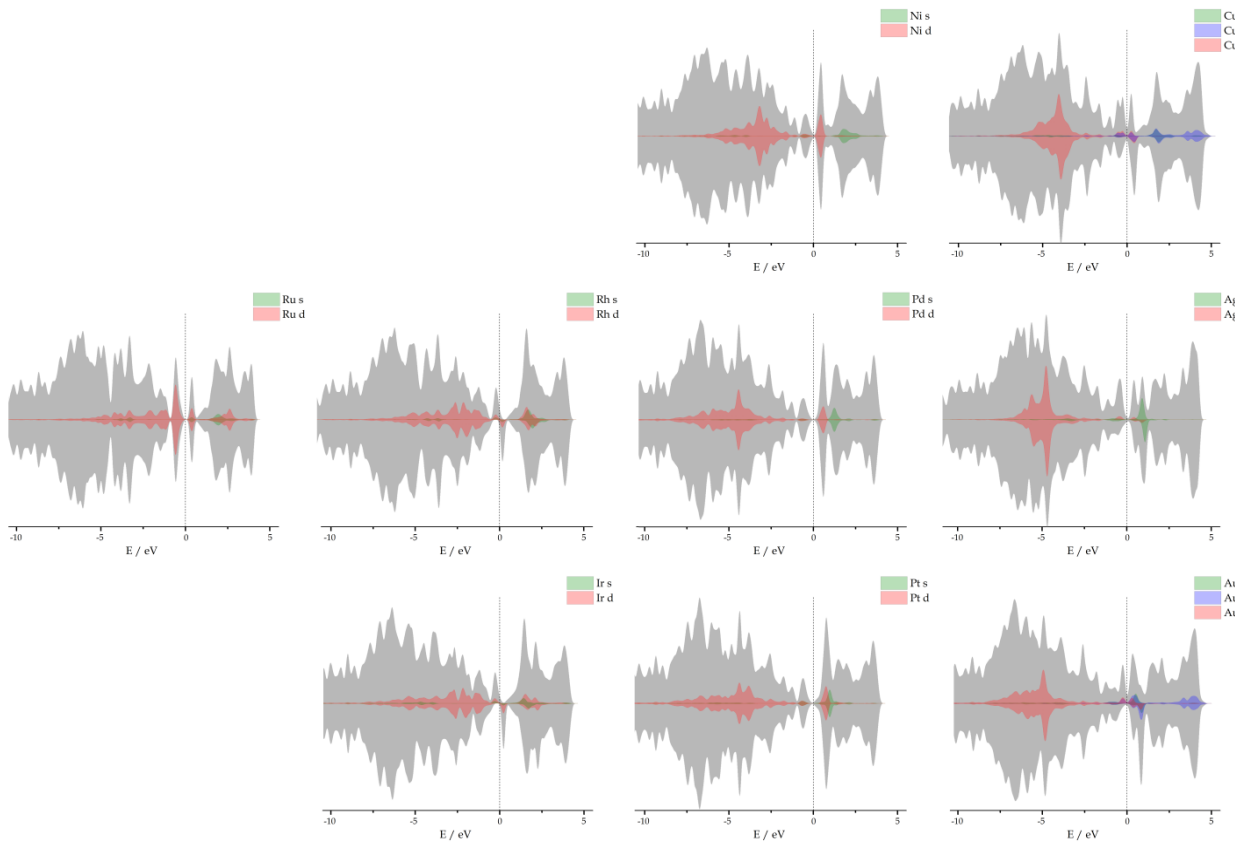

**Figure S4.** DOS plots for single atoms adsorbed on graphene with a single vacancy. The energy scale is referred to the Fermi level (vertical dashed line).

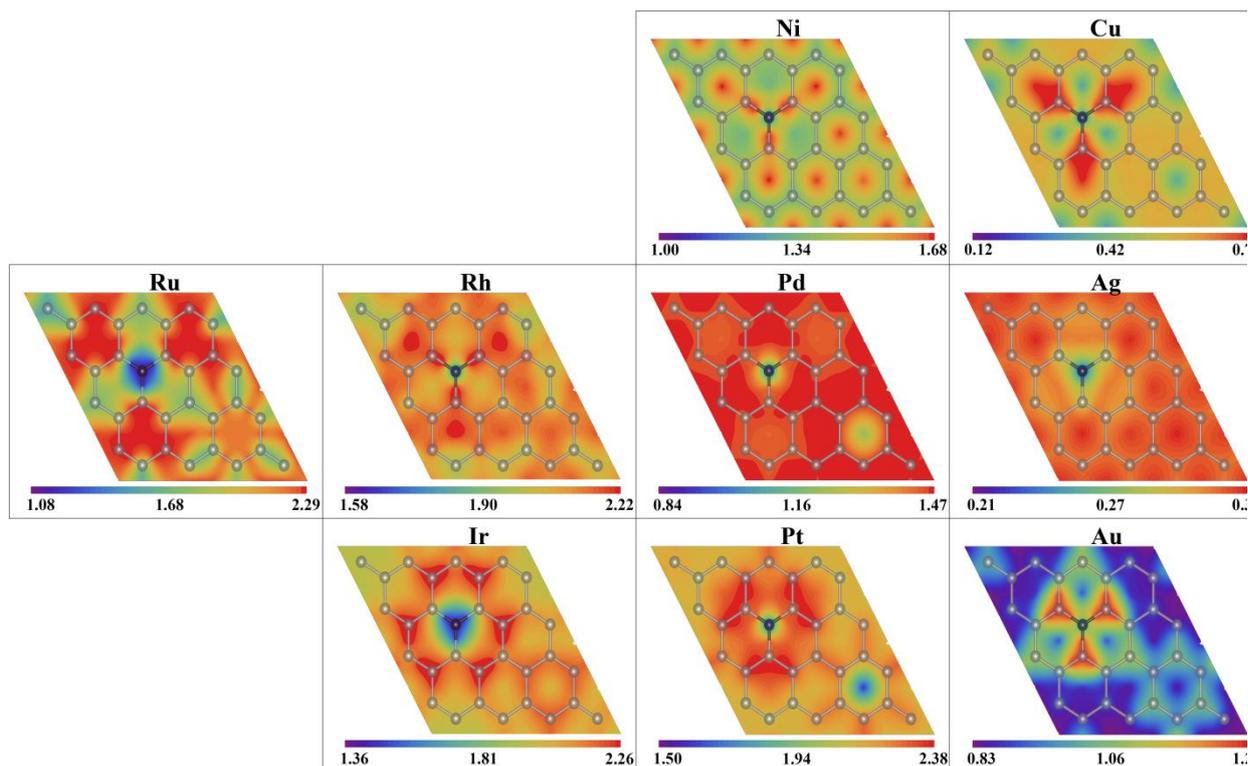

**Figure S5.** Heatmaps of $E_{\text{ads}}(M)$ in units of eV on NDG. Values in red represent stronger, while values in blue represent weaker bonding. Note that the scale is not the same for different systems.

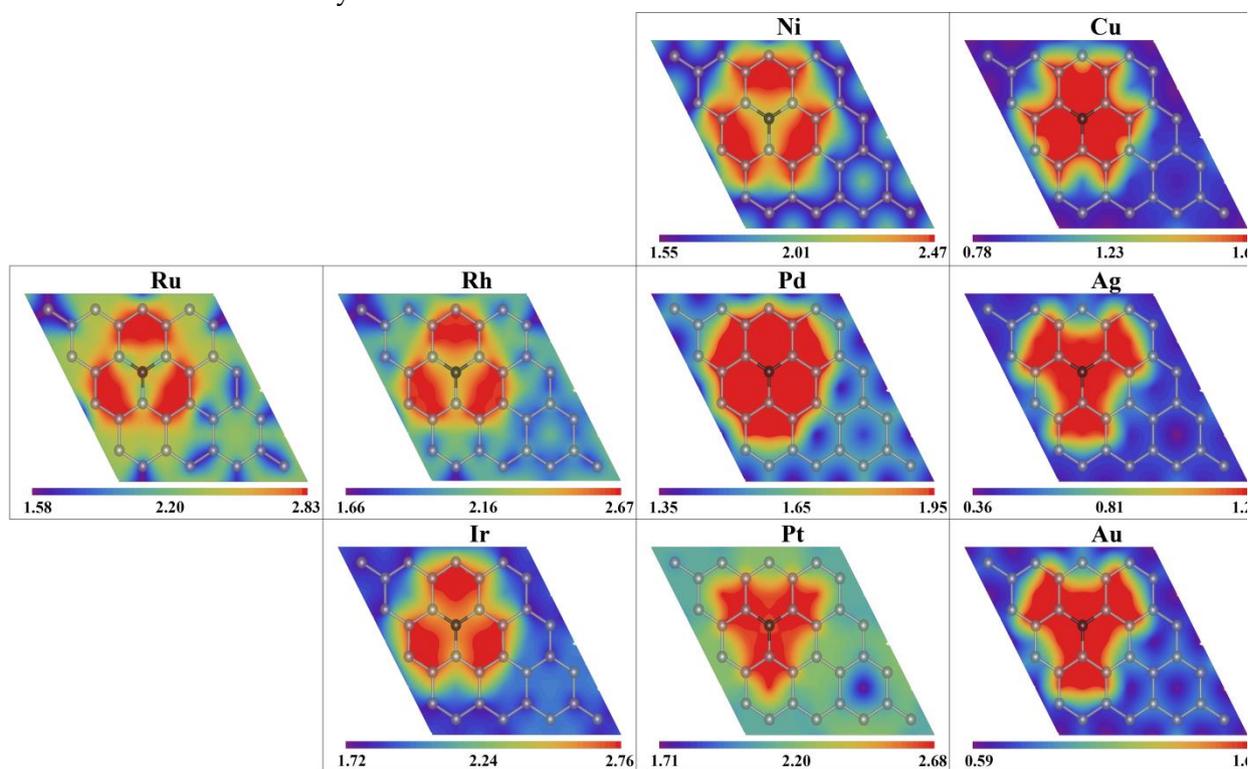

**Figure S6.** Heatmaps of $E_{\text{ads}}(M)$ in units of eV on BDG. Values in red represent stronger, while blue represents weaker bonding. Note that the scale is not the same for different systems.

**Table S1.** Calculated standard electrode potentials (*vs.* SHE) of $M^{z+}$/M@support couples ($E^o_{Mz+/M@support}$)

| M | z | $E^o_{Mz+/M}$ *vs.* SHE / V | $E^o_{Mz+/M@support}$ *vs.* SHE / V | | | |
|---|---|---|---|---|---|---|
| | | | Pristine | BDG | NDG | VG |
| **Ni** | 2 | −0.25 | −1.62 | −1.24 | −1.63 | 0.92 |
| **Cu** | 2 | 0.337 | −1.14 | −0.57 | −1.04 | 0.54 |
| **Ru** | 3 | 0.6 | −0.99 | −0.70 | −0.88 | 1.24 |
| **Rh** | 3 | 0.76 | −0.56 | −0.27 | −0.42 | 1.52 |
| **Pd** | 2 | 0.915 | −0.28 | −0.06 | −0.29 | 1.70 |
| **Ag** | 1 | 0.7996 | −1.83 | −0.89 | −1.82 | −0.11 |
| **Ir** | - | no data | - | - | - | - |
| **Pt** | 2 | 1.188 | −0.73 | −0.39 | −0.54 | 1.97 |
| **Au** | 3 | 1.52 | 0.48 | 0.81 | 0.68 | 1.19 |

**Table S2.** Adsorption energies of H, A (A = C, N, O and S) on M@pristine graphene model SACs.

|     | H     | X     | XH    | XH$_2$ | XH$_3$ | XH$_4$ |
|-----|-------|-------|-------|--------|--------|--------|
|     |       |       | C     |        |        |        |
| Ni  | −2.51 | −3.95 | −4.50 | −4.27  | −2.25  | −0.85  |
| Cu  | −3.37 | −3.41 | −4.32 | −4.51  | −3.09  | −0.06  |
| Ru  | −2.21 | −6.74 | −5.68 | −4.37  | −2.04  | −0.29  |
| Rh  | −1.94 | −5.55 | −5.43 | −4.44  | −2.38  | −0.31  |
| Pd  | −1.97 | −3.26 | −3.52 | −3.49  | −1.64  | −0.35  |
| Ag  | −2.58 | −2.30 | −3.02 | −3.37  | −2.17  | −0.07  |
| Ir  | −3.03 | −6.43 | −7.45 | −5.25  | −2.81  | −1.00  |
| Pt  | −2.76 | −5.39 | −4.91 | −4.91  | −2.41  | −0.99  |
| Au  | −3.53 | −3.70 | −4.08 | −4.45  | −3.09  | −0.11  |
|     |       |       | N     |        |        |        |
| Ni  |       | −2.72 | −3.26 | −3.12  | −1.62  |        |
| Cu  |       | −2.80 | −3.13 | −3.50  | −1.30  |        |
| Ru  |       | −4.56 | −4.19 | −3.25  | −1.39  |        |
| Rh  |       | −3.28 | −3.20 | −2.67  | −1.18  |        |
| Pd  |       | −2.04 | −2.10 | −2.15  | −1.31  |        |
| Ag  |       | −1.66 | −2.09 | −1.99  | −0.27  |        |
| Ir  |       | −4.76 | −4.99 | −3.49  | −2.00  |        |
| Pt  |       | −3.05 | −3.05 | −3.11  | −1.94  |        |
| Au  |       | −2.72 | −2.82 | −2.96  | −0.83  |        |
|     |       |       | O     |        |        |        |
| Ni  |       | −4.24 | −3.83 | −0.92  |        |        |
| Cu  |       | −3.90 | −4.43 | −0.62  |        |        |
| Ru  |       | −4.81 | −4.07 | −0.63  |        |        |
| Rh  |       | −3.98 | −3.61 | −0.60  |        |        |
| Pd  |       | −2.82 | −2.74 | −0.69  |        |        |
| Ag  |       | −2.34 | −3.32 | −0.07  |        |        |
| Ir  |       | −4.98 | −4.52 | −1.16  |        |        |
| Pt  |       | −4.00 | −3.56 | −1.05  |        |        |
| Au  |       | −3.26 | −3.64 | −0.24  |        |        |
|     |       |       | S     |        |        |        |
| Ni  |       | −3.92 | −3.38 | −1.68  |        |        |
| Cu  |       | −3.79 | −3.91 | −0.97  |        |        |
| Ru  |       | −4.64 | −3.55 | −1.12  |        |        |
| Rh  |       | −3.85 | −3.24 | −1.23  |        |        |
| Pd  |       | −2.83 | −2.53 | −1.35  |        |        |
| Ag  |       | −2.65 | −3.05 | −0.12  |        |        |
| Ir  |       | −4.67 | −4.12 | −2.14  |        |        |
| Pt  |       | −3.97 | −3.48 | −2.07  |        |        |
| Au  |       | −3.42 | −3.49 | −0.62  |        |        |

**Table S3.** Adsorption energies of H, A (A = C, N, O and S) on M@NDG model SACs.

| | H | X | XH | $XH_2$ | $XH_3$ | $XH_4$ |
|---|---|---|---|---|---|---|
| **C** | | | | | | |
| Ni | −2.79 | −4.02 | −4.61 | −4.34 | −2.44 | −0.84 |
| Cu | −3.07 | −3.12 | −4.10 | −4.21 | −2.80 | −0.09 |
| Ru | −2.83 | −6.46 | −5.58 | −5.07 | −2.55 | −0.31 |
| Rh | −2.58 | −5.14 | −5.12 | −4.31 | −2.20 | −0.36 |
| Pd | −2.22 | −3.38 | −3.73 | −3.66 | −1.83 | −0.58 |
| Ag | −2.73 | −2.30 | −3.23 | −3.53 | −2.28 | −0.07 |
| Ir | −3.24 | −6.83 | −7.13 | −5.52 | −2.93 | −0.95 |
| Pt | −3.30 | −5.06 | −4.79 | −4.74 | −2.89 | −0.60 |
| Au | −2.89 | −3.11 | −3.49 | −3.80 | −2.46 | −0.10 |
| **N** | | | | | | |
| Ni | | −2.77 | −3.32 | −3.60 | −1.59 | |
| Cu | | −2.51 | −2.84 | −3.18 | −1.17 | |
| Ru | | −4.35 | −4.08 | −3.29 | −1.36 | |
| Rh | | −2.98 | −3.05 | −2.63 | −1.27 | |
| Pd | | −2.07 | −2.21 | −2.52 | −1.30 | |
| Ag | | −1.82 | −1.44 | −2.20 | −0.61 | |
| Ir | | −4.74 | −4.02 | −3.72 | −2.01 | |
| Pt | | −2.91 | −3.20 | −3.37 | −1.66 | |
| Au | | −2.09 | −2.19 | −2.34 | −0.85 | |
| **O** | | | | | | |
| Ni | | −4.33 | −4.29 | −0.94 | | |
| Cu | | −3.72 | −4.26 | −0.65 | | |
| Ru | | −4.84 | −3.90 | −0.69 | | |
| Rh | | −3.75 | −3.46 | −0.60 | | |
| Pd | | −3.11 | −3.33 | −0.71 | | |
| Ag | | −2.77 | −3.34 | −0.14 | | |
| Ir | | −5.12 | −4.36 | −1.16 | | |
| Pt | | −4.10 | −4.00 | −0.85 | | |
| Au | | −2.80 | −3.15 | −0.35 | | |
| **S** | | | | | | |
| Ni | | −4.03 | −3.94 | −1.69 | | |
| Cu | | −3.58 | −3.72 | −0.79 | | |
| Ru | | −4.61 | −3.47 | −1.43 | | |
| Rh | | −3.76 | −3.16 | −1.33 | | |
| Pd | | −3.11 | −3.04 | −1.32 | | |
| Ag | | −2.88 | −3.04 | −0.18 | | |
| Ir | | −4.77 | −4.22 | −1.97 | | |
| Pt | | −4.04 | −3.80 | −1.72 | | |
| Au | | −2.99 | −2.94 | −0.60 | | |

**Table S4.** Adsorption energies of H, A (A = C, N, O and S) on M@VG model SACs.

|     | H     | X     | XH    | XH$_2$ | XH$_3$ | XH$_4$ |
|-----|-------|-------|-------|--------|--------|--------|
|     |       |       | **C** |        |        |        |
| Ni  | −2.40 | −4.55 | −4.93 | −4.23  | −2.19  | −0.76  |
| Cu  | −2.47 | −2.84 | −3.80 | −3.96  | −2.21  | −0.33  |
|     |       |       |       |        |        |        |
| Ru  | −2.81 | −6.36 | −5.62 | −4.93  | −2.70  | −0.25  |
| Rh  | −2.81 | −6.12 | −5.12 | −4.92  | −2.47  | −0.29  |
| Pd  | −2.34 | −3.99 | −3.79 | −3.37  | −1.50  | −0.54  |
| Ag  | −2.02 | −2.17 | −2.84 | −3.08  | −1.68  | −0.26  |
|     |       |       |       |        |        |        |
| Ir  | −3.58 | −6.99 | −6.82 | −6.12  | −3.31  | −0.39  |
| Pt  | −3.09 | −5.44 | −5.44 | −4.77  | −2.85  | −0.82  |
| Au  | −2.87 | −3.23 | −3.65 | −4.04  | −2.46  | −0.23  |
|     |       |       | **N** |        |        |        |
| Ni  |       | −2.90 | −3.31 | −3.01  | −1.56  |        |
| Cu  |       | −2.12 | −2.50 | −2.94  | −1.75  |        |
|     |       |       |       |        |        |        |
| Ru  |       | −5.00 | −3.96 | −3.64  | −1.57  |        |
| Rh  |       | −3.86 | −3.80 | −3.05  | −1.38  |        |
| Pd  |       | −2.14 | −2.36 | −1.97  | −1.26  |        |
| Ag  |       | −1.29 | −1.60 | −2.03  | −1.17  |        |
|     |       |       |       |        |        |        |
| Ir  |       | −5.48 | −5.34 | −4.11  | −1.80  |        |
| Pt  |       | −3.20 | −3.30 | −3.17  | −1.75  |        |
| Au  |       | −2.12 | −2.32 | −2.60  | −1.30  |        |
|     |       |       | **O** |        |        |        |
| Ni  |       | −4.09 | −3.87 | −0.92  |        |        |
| Cu  |       | −3.37 | −3.68 | −1.07  |        |        |
|     |       |       |       |        |        |        |
| Ru  |       | −5.21 | −4.43 | −1.02  |        |        |
| Rh  |       | −4.28 | −3.82 | −0.64  |        |        |
| Pd  |       | −2.79 | −2.81 | −0.72  |        |        |
| Ag  |       | −2.27 | −2.76 | −0.64  |        |        |
|     |       |       |       |        |        |        |
| Ir  |       | −6.05 | −4.58 | −1.05  |        |        |
| Pt  |       | −3.88 | −3.54 | −0.94  |        |        |
| Au  |       | −2.85 | −3.25 | −0.60  |        |        |
|     |       |       | **S** |        |        |        |
| Ni  |       | −3.73 | −3.24 | −1.52  |        |        |
| Cu  |       | −3.20 | −3.24 | −1.48  |        |        |
|     |       |       |       |        |        |        |
| Ru  |       | −4.62 | −3.32 | −1.42  |        |        |
| Rh  |       | −4.12 | −3.34 | −1.55  |        |        |
| Pd  |       | −2.72 | −2.34 | −1.28  |        |        |
| Ag  |       | −2.37 | −2.54 | −0.93  |        |        |
|     |       |       |       |        |        |        |
| Ir  |       | −5.73 | −4.30 | −2.15  |        |        |
| Pt  |       | −3.90 | −3.46 | −1.90  |        |        |
| Au  |       | −2.97 | −3.05 | −1.19  |        |        |

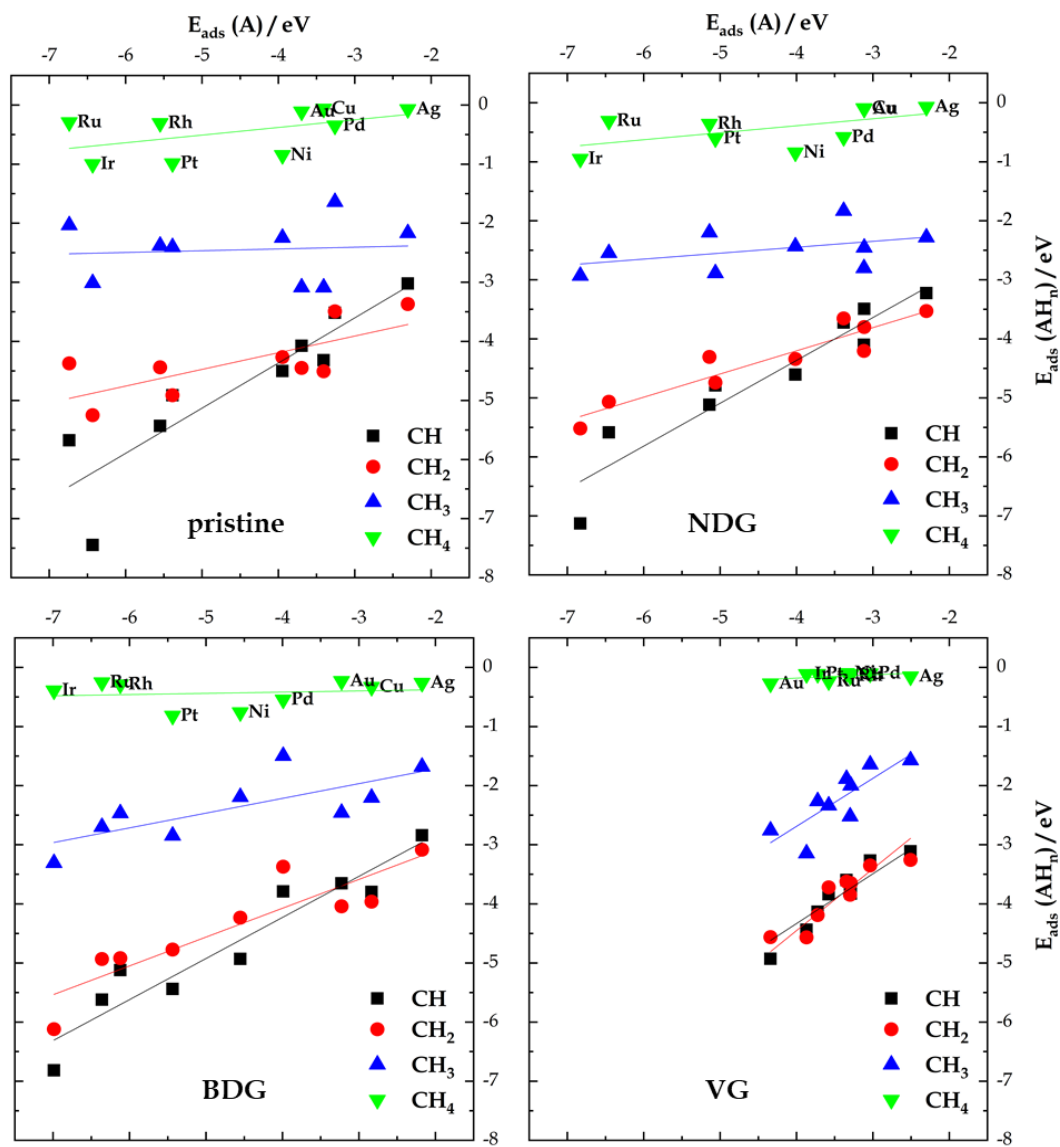

**Figure S7.** Scaling relations between $E_{ads}(C)$ and $E_{ads}(CH_x)$ ($x$ = 1, 2, 3, 4) for metal SAs on four studied supports

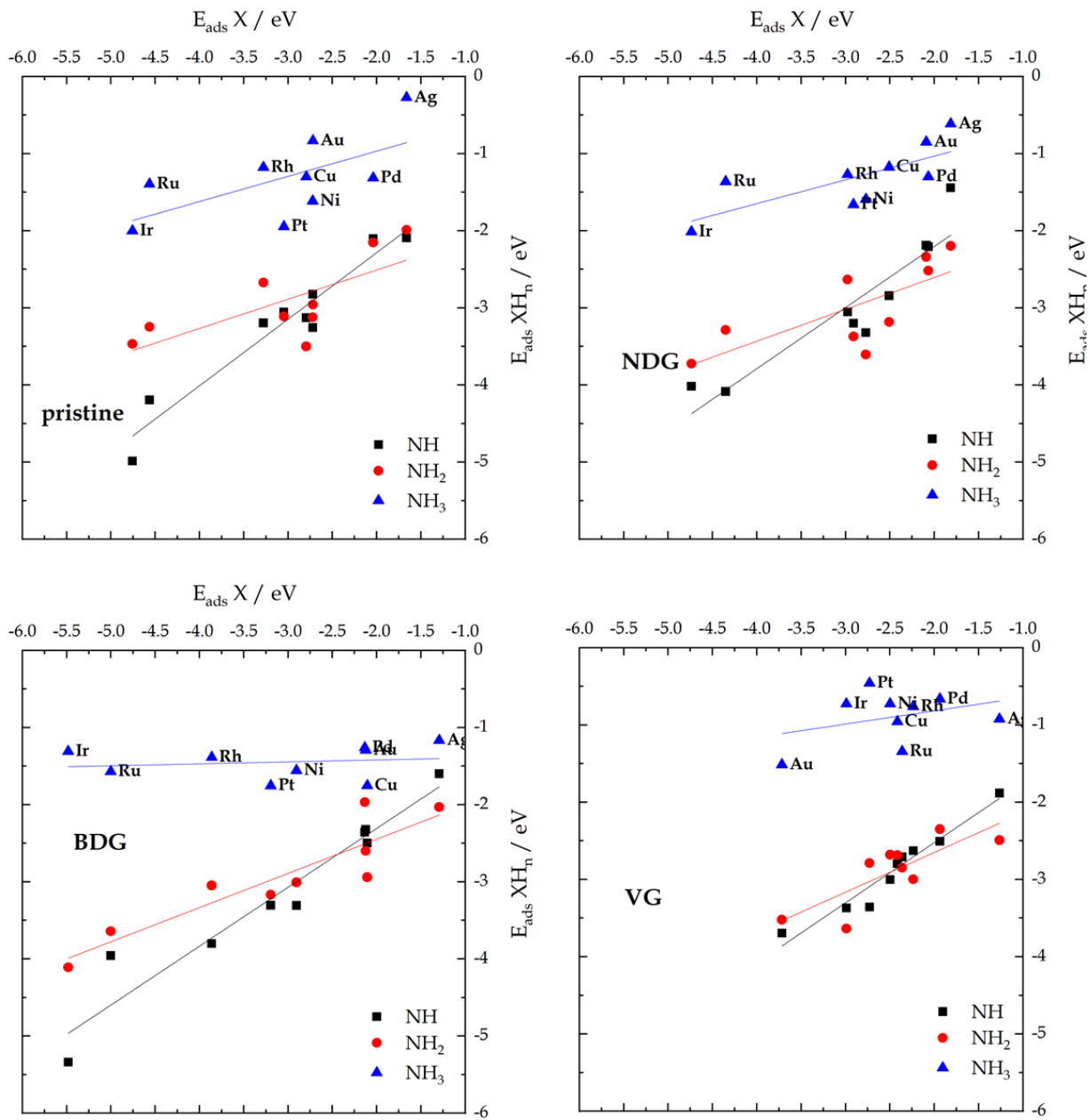

**Figure S8.** Scaling relations between $E_{ads}(N)$ and $E_{ads}(NH_x)$ ($x$ = 1, 2, 3) for metal SAs on four studied supports

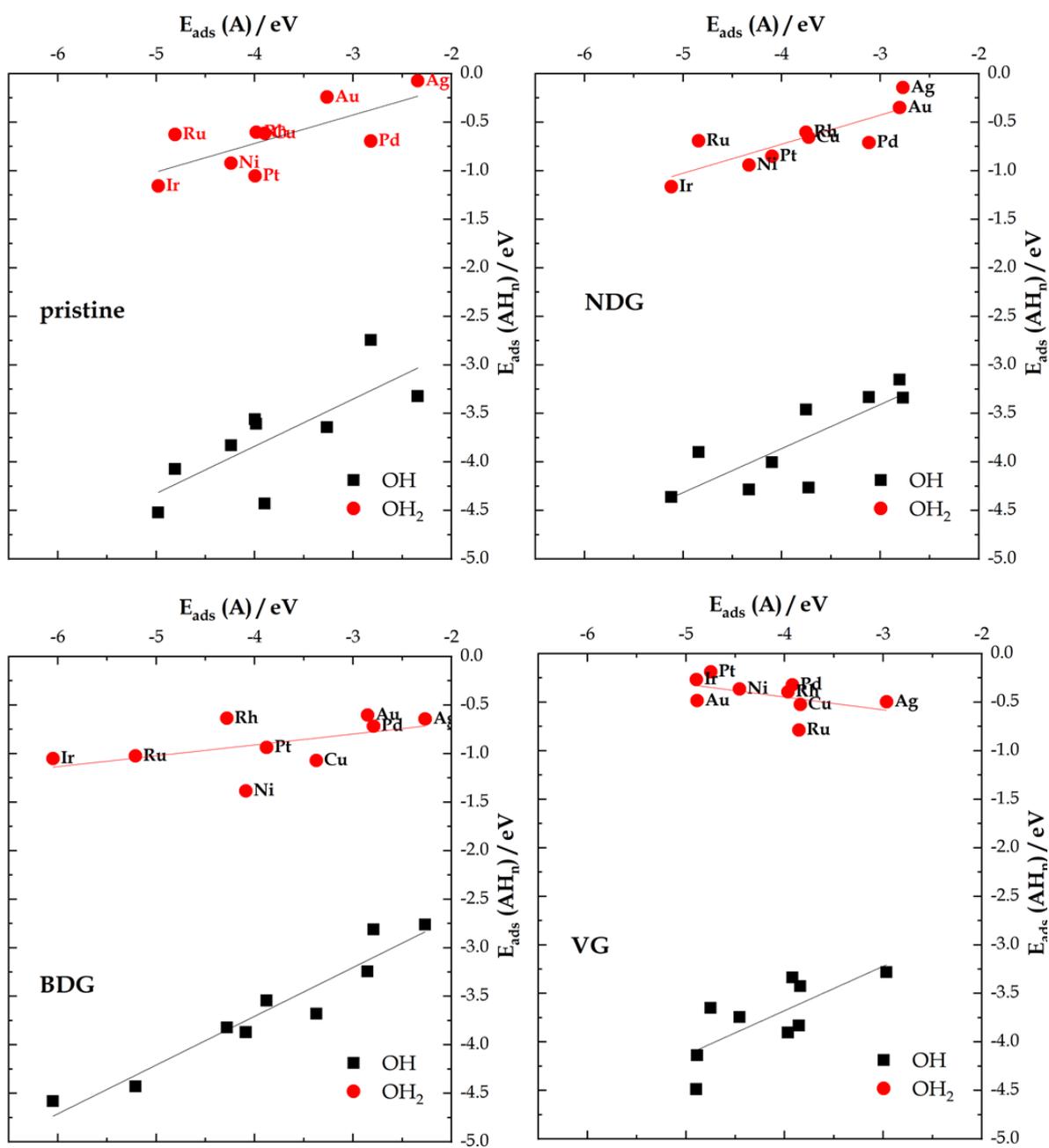

**Figure S9.** Scaling relations between $E_{ads}(O)$ and $E_{ads}(OH_x)$ ($x$ = 1, 2) for metal SAs on four studied supports

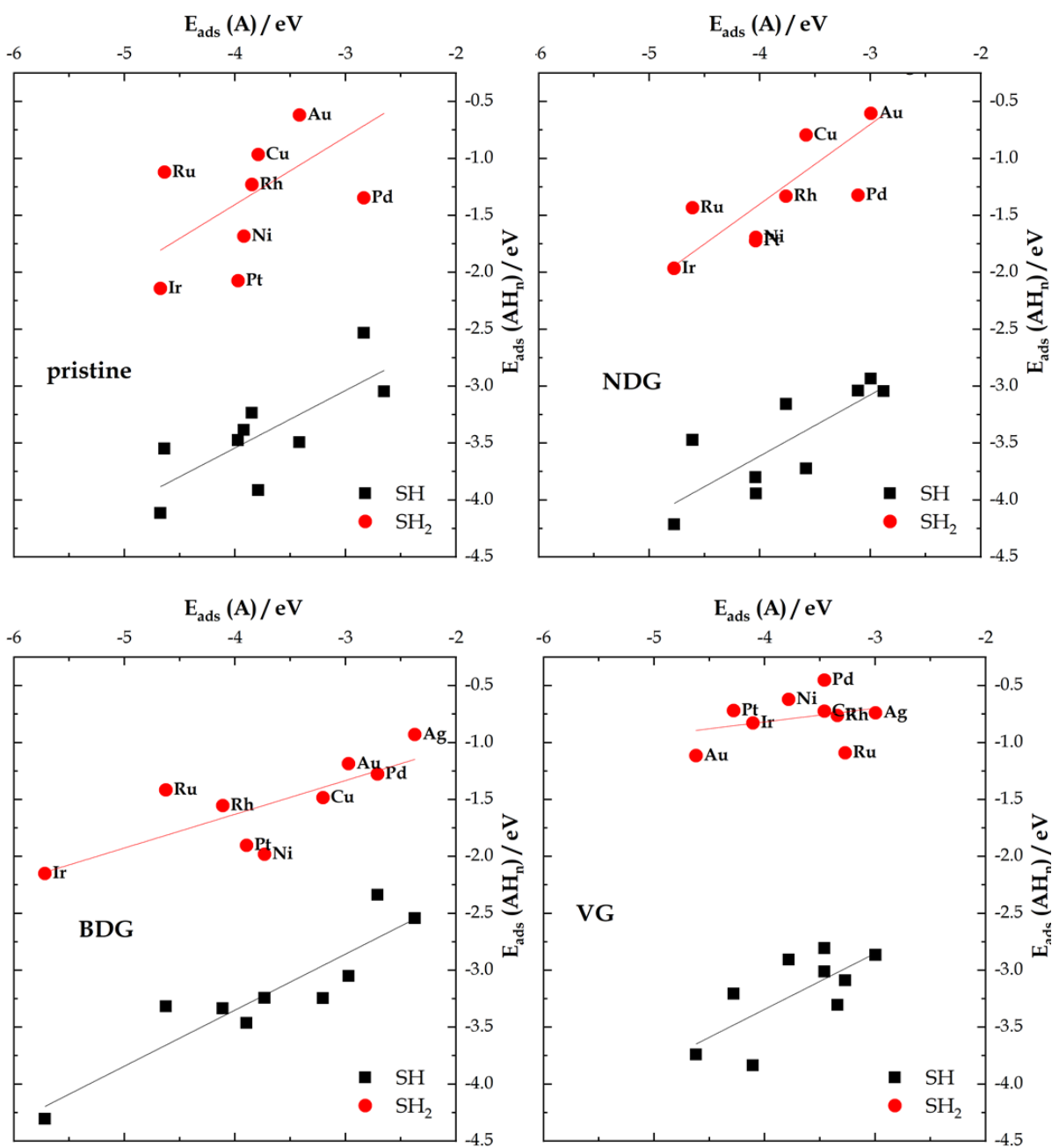

**Figure S10.** Scaling relations between $E_{ads}(S)$ and $E_{ads}(SH_x)$ ($x$ = 1, 2) for metal SAs on four studied supports

**Table S5.** Scaling relations for adsorbates when all studied supports are included along with the Pearson's correlation coefficient.

| Scaling | Intercept | Slope | Correlation coefficient |
|---|---|---|---|
| CH *vs.* C | -1.35 ± 0.22 | 0.738 ± 0.048 | 0.935 |
| CH2 *vs.* C | -2.48 ± 0.19 | 0.413 ± 0.042 | 0.862 |
| NH *vs.* N | -0.81 ± 0.14 | 0.773 ± 0.046 | 0.946 |
| NH2 *vs.* N | -1.75 ± 0.18 | 0.404 ± 0.059 | 0.808 |
| OH *vs.* O | -1.89 ± 0.21 | 0.471 ± 0.053 | 0.834 |
| SH *vs.* S | -1.43 ± 0.25 | 0.509 ± 0.067 | 0.795 |

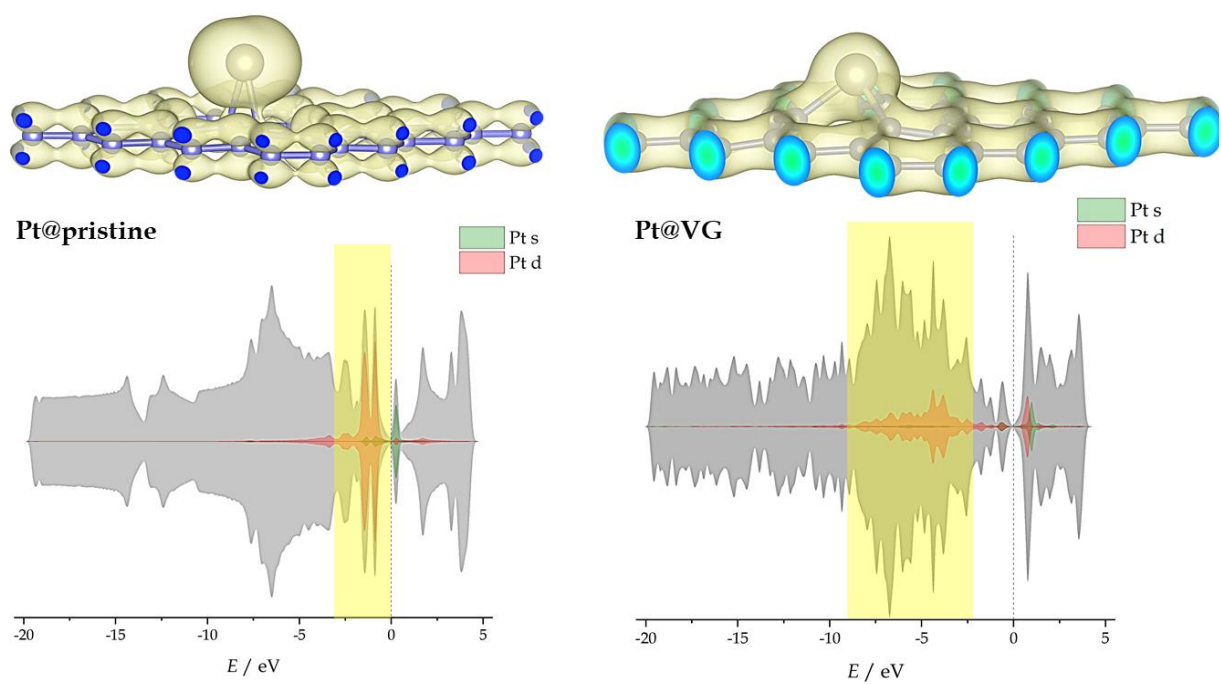

**Figure S11.** Upper row - integrated local density of states isosurfaces for Pt@pristine graphene (left) and Pt@VG (right). Lower row – density of states of the corresponding systems along with the indicated energy window in which the integration was performed.

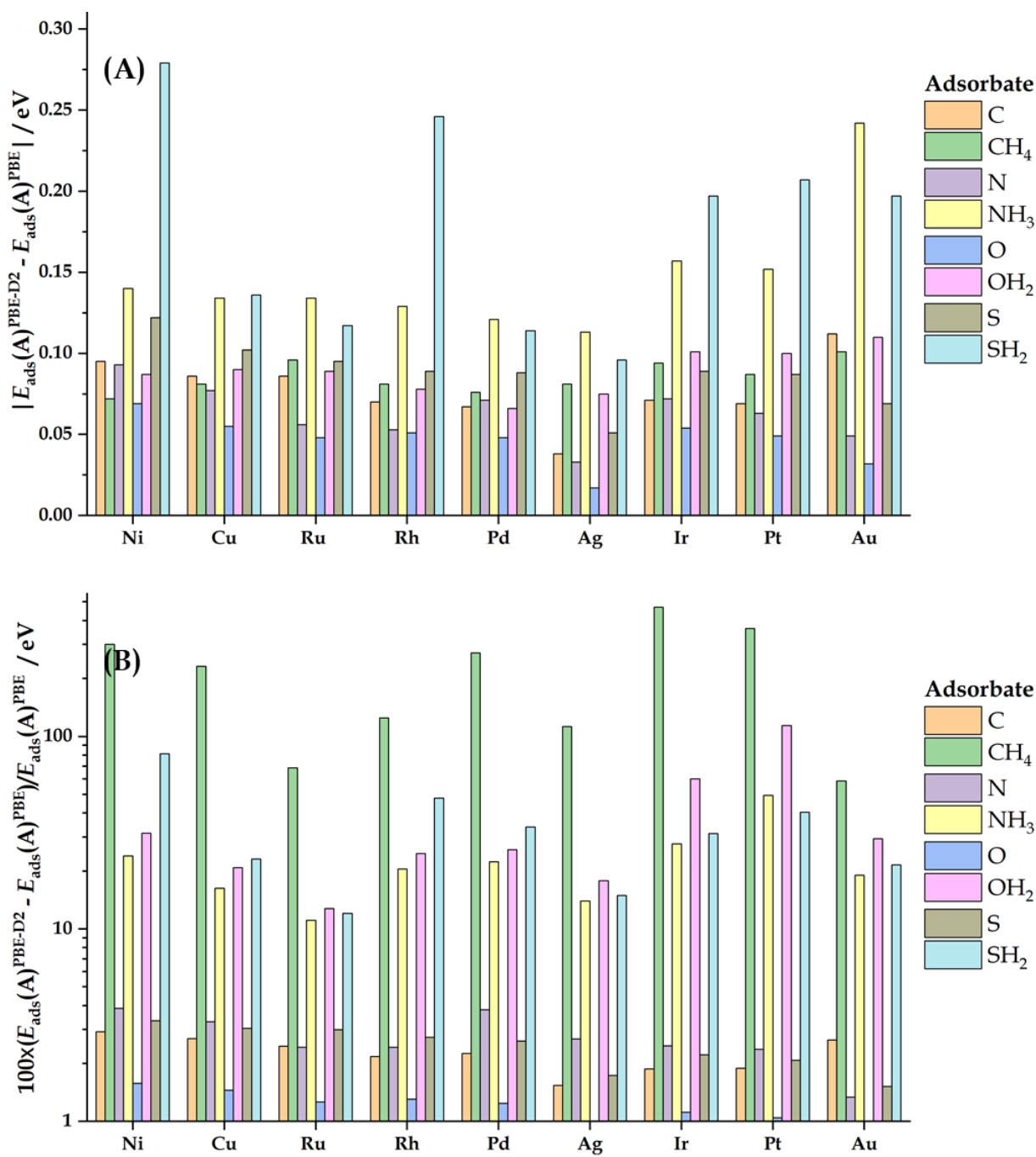

**Figure S12.** Absolute (A) and relative (B) contribution of dispersion interactions for studied adsorbates on M@VG model SACs.

**Table S6.** Hydrogen adsorption energies ($E_{ads}$(H)) at the C sites adjacent to the metal sites for the M@VG family.

| M@VG | $E_{ads}$(H) / eV | Total magnetization / $\mu_B$ |
|---|---|---|
| Ni | −2.10 | 0.84 |
| Cu | −2.52 | 0.00 |
| Ru | −1.63 | 1.00 |
| Rh | −2.04 | 0.00 |
| Pd | −1.75 | 0.31 |
| Ag | −3.18 | 0.00 |
| Ir | −1.87 | 0.00 |
| Pt | −1.57 | 0.03 |
| Au | −3.00 | 0.00 |